\pdfoutput=1
\documentclass[3p,twocolumn,a4paper,times,12pt,authoryear]{elsarticle}
\usepackage{graphicx}
\usepackage{amssymb}
\usepackage{wasysym}   % diameter&arrow symbols.

\usepackage{hyperref}

\begin{document}

\begin{frontmatter}
\title{The prolonged gamma ray enhancement and the~short radiation burst events observed in~thunderstorms at~Tien~Shan}

\author[lpi]{A.\,Shepetov}
 \ead{ashep@www.tien-shan.org}
\author[ion]{V.\,~Antonova}
\author[iet]{O.\,Kalikulov}
\author[ion]{O.\,~Kryakunova}
\author[e]{A Karashtin}
\author[ion]{V.\,~Lutsenko}
\author[lpi]{S.\,Mamina}
\author[iet]{K.\,Mukashev}
\author[lpi]{V.\,Piscal}
\author[lpi]{M.\,Ptitsyn}
\author[lpi]{V.\,Ryabov}
\author[iet]{N.\,Saduev}
\author[ipt]{T.\,Sadykov}
\author[ion]{N.\,Salikhov}
\author[f]{Yu.\,Shlyugaev}
\author[lpi]{L.\,Vil'danova}
\author[lpi]{V.\,Zhukov}

\address[lpi]{P.\,N.\,Lebedev Physical Institute of the Russian Academy of Sciences (LPI), Leninsky pr., 53, Moscow, Russia, 119991}
\address[ion]{Institute of Ionosphere, Kamenskoye plato, Almaty, Republic of Kazakhstan, 050020}
\address[iet]{Al-Farabi Kazakh National University, Institute of Experimental and Theoretical Physics, al-Farabi pr., 71, Almaty, Kazakhstan, 050040}
\address[e]{Radiophysical Research Institute, Bolshaya Pechyorskaya, str., 25/12a, Nizhny Novgorod, Russia}
\address[ipt]{Satbayev University, Institute of Physics and Technology,  Ibragimova str. 11, Almaty, Kazakhstan, 050032}
\address[f]{Institute of Applied Physics of RAS, Ul'yanova str., 46, Nizhny Novgorod, Russia}

\date{}

\begin{abstract}
We report the observation results of the hard radiation flashes which accompanied the lightning discharges above the mountains of Northern Tien Shan. Time series of the counting rate intensity, numerical estimations of  absolute flux, and energy distribution of accelerated electrons and of (20--2000)\,keV gamma rays were obtained at the height of 3700\,m a.\,s.\,l., immediately within thunderclouds, and in closest vicinity ($\lesssim$$100$\,m) to discharge region. Two different kinds of radiation emission events are presented here: a relatively prolonged rise of gamma ray intensity with minute-scale duration (the thunderstorm ground enhancement, TGE) which has preceded a negative field variation, and a short sub-millisecond radiation burst, which accompanied a close lightning discharge in thundercloud. It was revealed also an indication to positron generation in thunderclouds at the time of gamma ray emission, as well as modulation of the neutron counting rate in Tien~Shan neutron monitor which was operating at a (1.5--2)\,km order distance from the region of lightning development.
\end{abstract}


\end{frontmatter}

\section{Introduction}

By many experiments which were realized during two latter decades in the field of high energy atmospheric physics it was detected the phenomenon of a prolonged excess of the flux of high energy charged particles and local gamma radiation above its usual background which takes place in thunderstorm times. Typically, the relative amplitude of  the intensity increase seen during such events varies from a few tens to thousands of percents, and their duration may last from a few seconds order times up to tens of minutes. The energy range of gamma rays registered in the times of enhanced emission was of about (0.1--10)\,MeV, and that of the charged particles---up to the (10--100)\,MeV order. In many cases such radiation enhancements were immediately preceding, but not coinciding with, a subsequent lightning, and terminated just at the moment of a nearby lightning discharge. First time observed from an airplane by \cite{thunder_avia_first_1981}, later on the events of such type were multiple times detected in upper atmosphere both above thundercloud tops, e.g. by \cite{thunder_tgf_1985,kelly2015_thunderairborn,kochkin_inflighttge_2017,oestgard2019a}, and immediately within thunderclouds, with balloon-born detectors such as of \cite{thundermarshall3,eack2000}. In ground-based experiments these effects were observed at many particles detector installations situated at mountain altitudes, as by \cite{thundermounts3,aragatstge2013,thunderaragats2019,thunderneutrotibet2012,kudela_tge_stit_2017}, and at the sea level: \cite{thundertgejapan2002,thundertgejapan2011,tsuchiya2007,tsuchia_tge_2013,thundertge_bogomolov2015,thundertge_bogomolov2016,thundertgejapan2019}. In scientific literature a somewhat diverse terminology has been established for these phenomena which were referred to as ``prolonged radiation bursts'' in \cite{tsuchiya2007}, ``gamma ray glows'' by \cite{kochkin_inflighttge_2017}, ``thunderstorm ground enhancements'' (TGE) according to \cite{thundermounts3}. Hereafter in present publication the latter designation will be used, mostly because of its convenient shortness, as well as the fact that the subject of this paper is indeed connected with the data obtained in a ground-based experiment.

% classic runaway
Presently it is generally assumed that the appearance of the flow of high energy electrons in thunderstorm time is due to the effect of runaway electron breakdown---an acceleration of the charged particles in specific conditions when the energy acquired under the influence of an external electric field by some particle from the tail of their initial energy distribution (a ``runaway'') exceeds the losses of its interaction with surrounding matter. In turn, emission of gamma radiation in thunderstorm time is due to bremsstrahlung of accelerated electrons, as it was illustrated by many simulations, \textit{e.\,g.}, of \cite{thundermounts1,thundermounts4,babich_cascadesimu_2013,thundersimul2018}. An excessive neutron flux which has been detected in some most energetic electric discharge events within thunderclouds, such as observed by \cite{thunderneutrotibet2012,thunderneutroindia2015, thunderneutro2017}, may result from the photonuclear reactions caused by gamma rays: \cite{babich2007,thundermounts5,babich2013,babich2014,thunderpositron_enoto2017,bowers2017_thunderneutrojapan}. %,thunderneutro_babich_2019
Crucial role in formation of TGE events observed in ground-based experiments plays the lower region of positive charge distribution in thunderclouds which sometimes can appear below the main negative charge concentrated in the middle of the cloud, and ensures acceleration of electrons in direction to the earth's surface, as it was discussed in \cite{thundermounts2,thundermounts7}.

A possibility of appropriate conditions to exist inside thunderclouds for realization of the runaway breakdown mechanism has been predicted already by \cite{thunderneutrowilson}. Later on, this idea was quantitatively developed in a number of theoretical studies, such as \cite{gurmilrou,gurrou,thundersimul0,gurzyb}. Key statement of this theory is that a runaway electron when being accelerated by internal electric field of a thundercloud may produce another free electrons by its collisions with atmospheric atoms, and the latter may accelerate in their turn producing the next generation of the charged particles. As a result, a single seed electron put inside the region of the electric field with sufficiently high strength $E>E_{c}$ gives rise to a whole avalanche of high energy particles, consisting both of electrons and bremsstrahlung gamma ray quanta. The critical field $E_{c}$ is equal to 280\,kV/m at the sea level and diminishes with height proportionally to air density; this value is almost an order of magnitude below the threshold of the conventional dielectric breakdown in the air. As many rocket- and balloon-born experiments on direct sounding of thundercloud internals have shown, such as \textit{e.\,g.} those of \cite{thundermarshall1,thundermarshall2,thundersounding_stolzenbegr_2007}, generally the lightning discharges in thunderclouds were initiated when the strength of the electric field was close to $E_{c}$ value, while the fields essentially above $E_{c}$ were never detected. This experimental fact is an indirect evidence in favor of the runaway breakdown mechanism.

Next obligatory condition for beginning of avalanche process is the presence of the charged seed particles with sufficiently high, \mbox{$\gtrsim$(0.1--1)\,MeV}, initial energy inside the volume with critical field.
% As such, there were suggested the secondary electrons produced by cosmic rays in the atmosphere
%, as by \cite{stozhkov_cr_role_2003,lindy_2018_mean_secondary_cr_energy},
As such, the secondary electrons produced by cosmic rays in the atmosphere and $\beta$-electrons stemmed from natural radioactivity were suggested still by \cite{thunderneutrowilson}. The need of high-energy (relativistic) seed particles stipulated the commonly used term for designation of this model---the relativistic runaway electrons avalanche (RREA). As  well, for realization of such process it is necessary a sufficiently large geometrical dimension of the field region which is limited from below by characteristic size of developing avalanche, and must be of at least a hundreds of meters order; this condition is generally satisfied just in thunderclouds.

As it was shown by a number of simulations, such as  \cite{dwyer_2003_feedback,coleman_dwyer_2006_rrea_sims,roussel_2008_rrea_sims}, energy spectrum of gamma rays from an RREA generally is of an exponential form with the average energy of about 7\,MeV. The fact, that the spectra of gamma-emissions detected in thunderclouds sometimes were consistent with this result is another adequacy confirmation of the suggested theory: \cite{thundermounts1,tsuchiya2007,babich_2010_spectra_consistent_with_rrea}.

% thermal
Different type of the runaway particles based breakdown is the so called thermal, or cold breakdown process which is possible in a very high field, $E \gtrsim 3\cdot 10^3$\,kV/m. In thunderclouds, the fields of such strength can exist only transiently, and its distribution is limited by closest vicinity of the streamer tips in developing electric discharge leader. As a consequence, in contrast to RREA, in this case the development of electron avalanche takes place at much smaller distances of a tens of centimeters or a meters order scale. Another principal difference of the cold breakdown process from RREA is the absence of the need in external source of seed particles, since any free electron, including the ``cold'', or thermal-energy ones, can start avalanche acceleration in such strong fields. Later on, and under favorable conditions if a sufficiently high large-scaled field is present in the cloud, electrons accelerated due to the thermal process may play the role of seed particles for ``common'' RREA.

The model of the cold runaway breakdown was initially suggested by  \cite{gur1961_thermal_breakdown} and further on especially developed in application to thundercloud conditions in publications of \cite{dwyer_2004b_stepped_leaders_as_source,dwyer_2005b,moss2006_on_the_leader_tips,celestin2011_on_tips}. Multiple observations of the short, micro- and millisecond long intensive bursts of gamma radiation which were immediately coinciding with the moments of lightning discharges, such as reported by \cite{moore_2001_thundermounts_short_flashes,dwyer2004a_rocket_triggered,dwyer_2005a_stepped_leaders_as_source,montania_2014_thundermounts_short_flashes}, can be considered as an experimental prove of the model of particle acceleration at the streamer tips of developing lightning leaders.

% feedback
Essential modification added by \cite{dwyer_2003_feedback} to the runaway breakdown model is the relativistic feedback effect which consists of a surplus amplification of hard emission due to high energy  quanta of bremsstrahlung radiation and to positrons born through the $e^\pm$-pair production mechanism. Inside thundercloud field the positron component accelerates in opposite direction in relation to the electron flux, and produces on its way ionization electrons which occur to be additional seed particles for subsequent RREAs which generate another positrons, \textit{etc}. As a result, the multiplicity of both electrons and positrons grows exponentially which leads to a significant, up to an order of magnitude, increase of overall gamma-emission from the discharge region comparatively to single RREA. As suggested by \cite{dwyer_2008_feedback_tgf}, it is the runaway discharge process in combination with positive relativistic feedback which lays in the basis of the mostly fast and energetic phenomena of the high energy atmospheric physics---terrestrial gamma ray flashes (TGF), such as seen by \cite{thunderorbits0,thunderorbits1,thunderorbits2,thunderorbits3} in upper atmosphere, and of analogous downward TGFs which have been observed in a number of ground-based experiments, such as of \cite{thunder_downtgf_dwyer_2012,thunder_downtgf_2015,thunder_downtgf_2016,wada2019a}. On the other hand, an equilibrium discharge current arising because of steady generation of RREAs amplified by positive feedback may compensate continuous charge separation in thundercloud and ensure a comparatively prolonged quasi-static state which reveals itself for outer observer as a long lasting gamma ray glow or a TGE event.

A comprehensive survey of modern progress in the theory of runaway electrons acceleration and its implication to the problems of atmospheric electricity can be found in the review of \cite{dwyer_review2012}.

% mos
An alternative mechanism which could be responsible for detection of an additional flux of high energy particles in thunderstorm times, and primarily for appearance of long-lasting TGE events, is the modification effect of the energy spectrum (MOS) of charged particles background under the influence of thundercloud field. According to this model which has been initially suggested in
\cite{aragatstge2011b,thunderaragats2013_gamma_spectra, thunderaragats2014_on_mos_again},
by favorable orientation of atmospheric electric field both electrons and positrons originated from cosmic rays interaction can acquire an additional energy which leads to the rise of their lifetime and attenuation length in atmosphere, and to corresponding intensity increase of detected radiation at the observation level. With account to MOS effect an appearance of the atmospheric electricity connected phenomena is possible even in those cases when the thundercloud field does not exceed the $E_{c}$ limit. As it was illustrated in \cite{thundersimul2017_mos, thunderaragats_2017,thunderaragats2018}, detecting of TGE events in combination with an intensive simulation of particles interaction in atmospheric fields can be a mean to probe the structure of electric field distribution in thunderclouds.

% radon
One more source of the excessive flux of high energy charged particles and gamma rays in thunderstorm times may by the decay of radioactive nuclei, primarily the radon $^{222}$Ra isotope and its daughter products whose concentration increases in near-earth atmosphere because of precipitations which usually accompany the periods of thunderstorm activity. As it is often stated, \textit{e.\,g.} by \cite{thunderaragats2018_with_radon}, such a temporary rise of the local radioactive background can be accountable for the most prolonged (tens of minutes) and low-energy (below 1\,MeV) part of TGE events.

Thus, thanks to intensive efforts undertaken in the course of last decades it was found a variety of unexpected phenomena in such a manifold investigation field as the high-energy atmospheric physics, and a number of theoretical models was created for their explanation. In spite of this progress the general picture of atmospheric electric discharge, and primarily the processes which lay in the basis of lightning initiation, still remain not completely clear. This circumstance necessitates further \textit{in situ} observations and collection of as much as possible new experimental data on radiations which accompany electric discharges in thunderclouds.

% on station
An appropriate place for such investigations is Tien Shan Mountain Cosmic Ray Station of Lebedev Physical Institute. At multiple studies executed here in 2000s a number of observations were made of the intensive gamma rays and high energy electrons bursts at thunderstorm time, some of them being similar to what was designated later on as TGE and TGF in literature: \cite{thunderour1999,thunderour2003,thunderour2009,thunderour2004,thunderour2009c,thunderour2011,thunderour2013}. Since then a special complex of detector facilities was created at Tien Shan station which is especially aimed to experimental investigations in the range of high energy atmospheric physics \cite{thunderour2016,thunderour2018opti}. A key feature of this complex is its ability to detect the soft gamma radiation and accelerated charged particles immediately within thunderclouds, at a small distance to development region of lightning discharges. For this purpose a special high altitude detector point was created in the neighborhood of Tien~Shan station which is placed on the top of a mountain ridge, at an altitude of 3700\,m above the sea level, and $\sim$400\,m above the average height of the station. In the time of thunderstorms this point frequently occurs to be deeply immersed within thundercloud, so detection of various radiations is possible at a few tens and hundreds of meters order distance from the region of their generation in close lightning discharges. Such disposition permits to avoid any significant influence on the part of the scattering and absorption processes, and to expand the range of detected radiations into the low-energy part of their spectrum. As a result, the lowest energy threshold in the Tien~Shan experiment now is of about (20--30)\,keV for gamma rays, and about (1--2)\,MeV for accelerated electrons. This is an essential difference from, correspondingly, a few hundreds of keV and tens of MeV detection thresholds of gamma rays and electrons which were typical for the measurements made previously in other ground-based experiments, at a kilometer order distances from thundercloud.

The subject of current publication is to present the results  obtained by observation at the high altitude detector point of Tien Shan station of the two close events observed in the times of thunderstorm activity: a TGE type prolonged gamma ray glow which preceded a close atmospheric discharge, and a hard radiation burst from a nearby lightning. In both cases the distances between the measurement point of radiation intensity and the region of lightning development occurred of the $\lesssim$100\,m order only.

\section{Instrumentation}

\subsection{The detector of gamma radiation}

For registration of the soft gamma rays which accompany the lightning discharges at Tien Shan station it is used a scintillation detector on the basis of a cylindrical \diameter 110 $\times$ 110\,mm$^2$ NaI(Tl) crystal coupled with a photomultiplier tube (PMT). Since the amplitude of any particular electric pulse at PMT output is proportional to the energy of corresponding gamma ray quantum absorbed in scintillator (supposing the conversion linearity of scintillation flash into electric signal), an amplitude analysis of output pulses permits to recover the energy spectrum of detected radiation. For this purpose the signals from the detector output come to a set of analog pulse discriminators with consecutively increasing operation thresholds. Every time when the amplitude of scintillation light exceeds the given threshold, the corresponding discriminator channel generates a single standard pulse signal which can by counted by a digital scaler scheme. Fixed length of these signals is set to 10\,$\mu$s which is compatible with the own decay time of used scintillator.

An absolute energy calibration of the whole amplitude measurement system was made with a set of radioactive gamma sources.

Concrete values of the radiation detection thresholds were somewhat different in the course of considered experiment. In the measurement season of the year 2018, when it was detected the TGE event which will be presented below, there were 14~amplitude channels with their thresholds distributed between $\sim$20\,keV and 2.5\,MeV. At the time of the short radiation burst which happened in summer 2017 there were only 12~channels with an upper energy threshold of 1\,MeV.
%   1  2   3  4  5   6   7   8   9  10  11   12  13   14
% 13.7.2017:
% 27 30 100 150 200 300 450 600 650 700 850 1000 (total 12)
% 12.8.2018:
% 20 25  30  70 100 250 350 500 600 700 800 1600 2000 2400
%                                                (total 14)

\begin{figure}
\centering
\includegraphics[width=0.49\textwidth, trim=0mm 0mm 0mm 0mm]{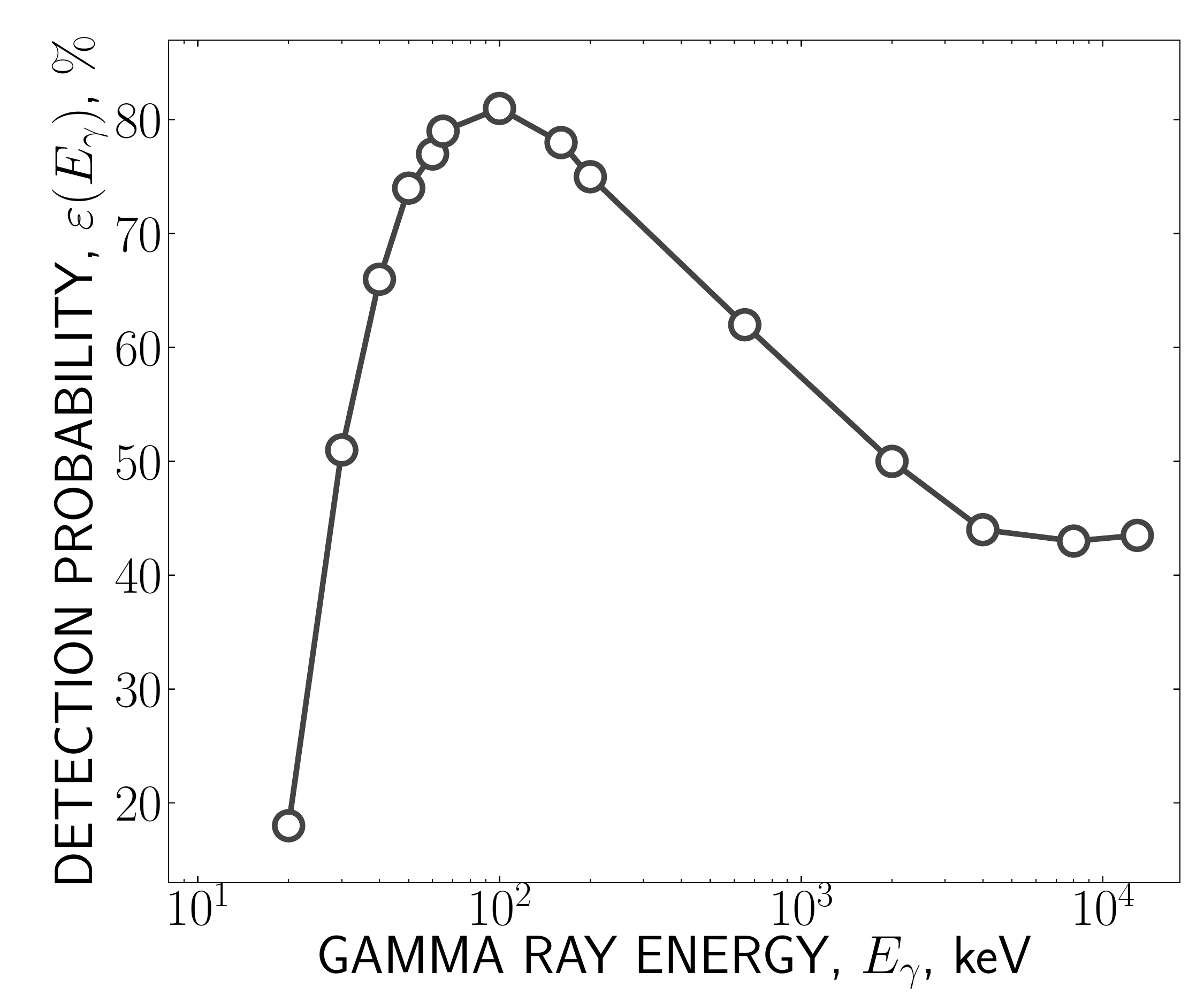}
\caption{Probability of gamma quantum registration (detection efficiency) of the gamma radiation detector.}
\label{figigammoeffici}
\end{figure}

The registration probability of gamma radiation in dependence on its energy, \textit{i.\,e.} the efficiency of the considered detector was obtained through a Geant4 based simulation of the gamma ray interaction processes with internal material of scintillator. In this calculation the absorption probability $\varepsilon$ of a gamma ray quantum within the scintillator was defined for a number of the fixed values of radiation energy $E_\gamma$. Since neither the probability of subsequent scintillation light emission, nor the efficiency of PMT photocathode were taken into account by this simulation, the values  $\varepsilon$ thus defined should be considered in the sense of a possible upper limit of detection efficiency. These estimations are presented in Figure~\ref{figigammoeffici} in dependence on corresponding $E_\gamma$ values.

\subsection{Charged particles detector}
The detection of electrons accelerated in the electric field of thundercloud in Tien Shan experiment is realized by a coincidence telescope made of a pair of large sized 1$\times$1$\times$0.01\,m$^3$ charge particles detectors. Each detector consists of a set of the square shaped plates of molded plastic scintillator interlaced with the light conducting optic fibers, as explained in \cite{shalour2006}. According to the detector testing results presented in that publication, the detection probability of relativistic charged particles by a single scintillator of such a kind is of about 0.95--0.99, while that for the soft gamma rays is quite negligible ($\lesssim0.01$). Both these estimates were obtained experimentally in the measurements with the calibrated radioactive sources.

The time duration of the output pulse signals elaborated by shaper schemes of the charge particles detector are of about (2--3)\,$\mu$s.

In the telescope set-up used at the high altitude point of Tien Shan station both scintillators were installed horizontally one above the other inside a wooden cabin with the roof thickness of about 1\,g/cm$^2$, and with a 40\,g/cm$^2$ thick layer of additional absorbing material placed between them.  Hence, the signals registered from the upper scintillator of such detector correspond to relativistic electrons with the energy of at least 2--3\,MeV, and observation of coincidence between the pulses from the upper and lower scintillators means nearly vertical passage of an electron with the energy above 80--100\,MeV, in dependence on the slope of its trajectory.

% from 'remarks2016.txt':
%nominally, there are 12 rubber layers between plastic scintillation detectors in the point, but the mean absober thickness should be set to 7-8 layers.
% 2.5cm x 2.1 g/cm^3 = 5 g/cm^2
%                 x 8 = 40 g/cm^2
%          x 2.5 MeV / g/cm^2 = 90 MeV
% for Al housing:
%   0.1cm x 2.7 g/cm^3 x 2.5 MeV / g/cm^2 = 0.7 MeV
% for a 1 cm thick wooden roof:
%   1cm x 0.7 g/cm^3 x 2.5 MeV / g/cm^2 = 1.8 MeV
%                     resulting sum: 0.7+1.8=2.5 MeV

\subsection{High altitude detector point}

To avoid unwanted absorption of radiations on their way from generation region in thundercloud to detector, both the gamma rays and the charged particles detectors used in the discussed experiment were installed at a high altitude point which is situated just on a mountain top, 400\,m above the common level of Tien Shan station. Quite often during thunderstorms this point occurs to be deep in the clouds, and there were some cases when the lightning was developing at a distance of about a few tens of meters only from the detectors.

Special measures were taken to ensure stable operation of electronic hardware placed in such unordinary conditions. All necessary components both of the detector and of the data acquisition (DAQ) system installed there were made as compact as possible to have  minimum length of all connecting wires. At thunderstorm time any external cable lines were physically disconnected from the high altitude point, so the powering of its equipment was performed completely from an in-build battery source. This battery, the whole DAQ electronics, the data registration computer, and the gamma ray detector were installed together within a same $0.5\times 0.5\times 1$\,m$^3$ Faraday cage welded of a solid, 1\,mm thick iron sheet, and by such a way that neither internal wire connection between them exceeded the length of (0.2--0.3)\,m. Two scintillation signals of the charged particles telescope were connected to this DAQ system by a pair of shielded, 0.5\,m long coaxial cables.

Intensity registration of the digital pulse signals in the high altitude point is realized by a compact DAQ system with reduced power consumption which is built on the basis of a STM32 type microcontroller  \cite{our2017_stm32}. There are 17 separate signals connected to this system: 14 outputs of the gamma detector amplitude discriminator, the outputs of the upper and lower scintillator plates of the charged particles telescope, and the coincidence pulse between the latter two signals. For all these signals the microcontroller driver program ensures the measurement of their counting rates simultaneously in two modes. These are the regular monitoring mode of the average levels of signal intensity with periodicity of 1\,s, and the detailed registration of the time histories of signal development with two temporal resolutions of 160\,$\mu$s and 800\,$\mu$s.

The high resolution time series of signal intensity registered by the microcontroller DAQ system may be strictly bound to some specific event which in the case of the considered experiment typically coincides with the moment of lightning discharge. For this purpose it is used a special control signal---the trigger. This trigger can be either generated by a hardware sensor of the fast jump in the strength of local electric field in detector vicinity, or it can be elaborated internally by the microcontroller driver program itself. In latter variant the program looks continuously after behaviour of all input signals, and in the case if some logical condition is fulfilled (\textit{e.\,g.} if the current sum of the pulse numbers stored over a specified time period exceeds a predefined threshold limit) it fires an internal trigger signal which initiates recording of the next event. For the high altitude detector point where any outside connections are strictly prohibited at thunderstorm time, it is the internal logical type of trigger which is applied for synchronization of the detected gamma ray and electron flux series with the lightning moment.

Regardless of synchronization type, the DAQ system operates always in  pre-trigger/post-trigger registration mode: the current multiplicity values for all input signals measured with 160\,$\mu$s or 800\,$\mu$s time resolution are kept continuously in cyclic buffer within microcontroller memory, so that whenever any trigger comes, the whole history of signal development is available both before and after its moment. In the time of the discussed experiment the duration of both the pre- and post-trigger time series in every registered lightning event was accepted to be equal to 1\,s.

\subsection{Neutron detectors}

The data on the behaviour of the neutron flux intensity in thunderstorm times which will be discussed further on were obtained at the neutron detector site of the Tien Shan station, 400\,m below the thunderstorm radiations detector system of the high altitude point. The straight line-of-view distance between the latter and the neutron detectors is of about (1.5--2)\,km.

There are two detector kinds for the neutron flux measurements at Tien Shan Mountain Station. The first is a NM64 type neutron supermonitor, such as described by \cite{hatton_supermonitor,hatton_supermonitor_inbook}, which operates here continuously during many decades \cite{ontienmonitor}. Generally, the neutron monitors of such type can be used for a high precision registration of the intensity of energetic ($\gtrsim$1\,GeV) cosmic ray hadrons capable to generate a multitude of evaporation neutrons in nuclear interaction within the internal lead target of the monitor. Besides particles of the  cosmic ray hadronic component, the neutron monitor has a residual sensitivity to the local background of low-energy neutrons with detection probability of the order of $\lesssim$(0.5-1)\% only: \cite{clem2000,nm64effici_shibata2001,yanke2011}.

Together with cosmic ray hadrons, the neutron monitor may detect high energy muons which possibility was particularly analyzed by \cite{our2018_undg}. This  effect is due to the high energy brems\-strah\-lung radiation emitted at muon passages, with subsequent neutron production in photonuclear reaction of resulting gamma rays, to direct nuclear interactions of the muonic cosmic ray component, and to the $\mu^-$-capture mechanism. As well, the photonuclear interaction channel ensures detection of the high energy ($\gtrsim$10\,MeV) gamma rays, electrons and po\-sit\-rons by the monitor, thou with a probability $\sim$(10--30)~times below the efficiency of hadron registration: \cite{thunderneutrotibet2012,babich2007,babich2013,babich2014,thunderneutro2017,thunderneutro_babich_2019}.

Tien~Shan neutron monitor consists of three 2$\times$3\,m$^2$ units each of which includes six big ($\diameter$0.15$\times$2\,m$^{2}$) gas discharge counters. The counters are sensitive to low-energy (thermal) neutrons because of their special filling---the enriched $^{10}$BF$_3$ gas. To ensure the fast energy loss by evaporation neutrons originating from nuclear reactions down to thermal values, the neutron moderator layers of a light hydrogen reach material were included into monitor set-up. Registration of the shaped pulse signals from the anode wires of these counters is realized by a STM32 microcontroller DAQ system of the same type as what is used at the high altitude point. The periodicity of the neutron intensity measurements accepted at Tien~Shan monitor is one minute, which is a standard in the world wide net of the cosmic ray intensity monitoring.

Second kind of neutron measurements at Tien Shan station is fulfilled with the low-threshold detectors consisting of a set of ``nude'' neutron-sensitive counters which are used without any surrounding heavy target neither moderator material. Because of such configuration these counters are sensitive mostly to the flux of thermal neutrons born under the influence of cosmic rays in outer environment around the detector.

\begin{figure}
\centering
\includegraphics[width=0.49\textwidth, trim=0mm 0mm 0mm 0mm]{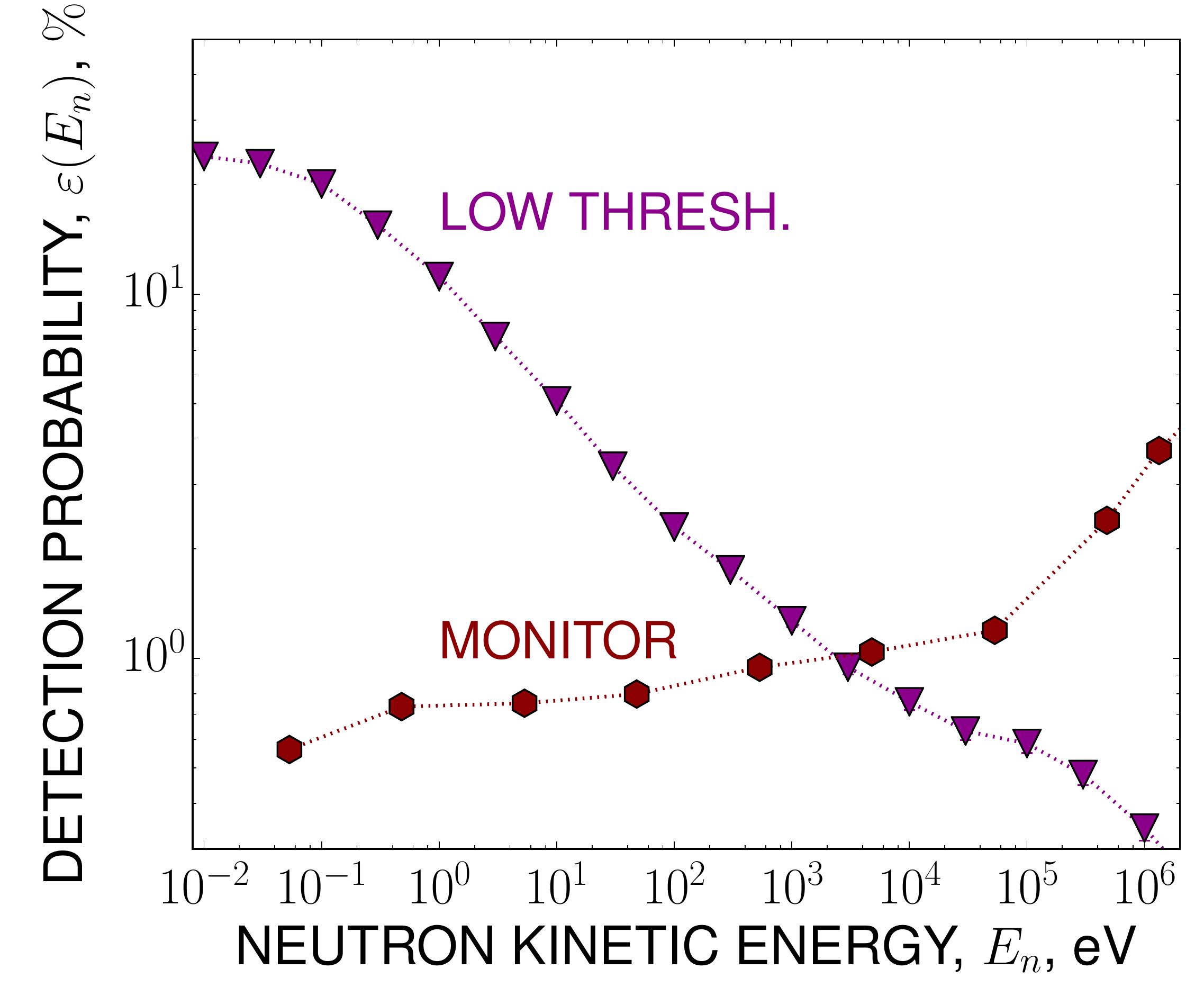}
\caption{Detection efficiency of the NM64 neutron supermonitor and of the low-threshold neutron detector consisting of 6~gas discharge neutron counters.}
\label{figineutroeffici}
\end{figure}

The neutron registration efficiency in the experiments of Tien Shan station was defined by a Geant4 based model simulation which took into account characteristic features both of the neutron detector internals, and of the outer environment typical for the station. %\cite{yanke2011}.
In Figure~\ref{figineutroeffici} it is presented the principal result of these simulations---the dependence of detection efficiency on the energy of incident neutron which was obtained for both detector types: for the NM64 neutron supermonitor and for the low-threshold neutron detector. Similar data on this subject were reported by \cite{clem2000} and \cite{nm64effici_shibata2001}.

\subsection{The electromagnetic emission, electric field, and acoustic signal detectors}

Temporal history of atmospheric discharge in the lightning events of Tien Shan experiment can be precisely traced by the records of electromagnetic emission which generally accompanies development of the electric streamers in thundercloud. At the time of considered experiment the signals of electromagnetic emission were detected simultaneously in two frequency ranges: MF/HF (0.1--10\,MHz), and VLF (1.5--11.5\,kHz). For this purpose a pair of corresponding antenna sets together with subsequent receiver electronics is now in use at Tien Shan station. The analog output signals of both receivers were operated by two amplitude-to-digital converter (ADC) systems. The digitization was made with a 12-bit accuracy, and with  periodicity of 0.16\,$\mu$s (in the MF/HF range) and $200$\,$\mu$s (for the VLF signal). The total duration of the time series records for both signals in detected events was of about (2.5--3)\,s.

Along with waveform recording, analog output of the MF/HF receiver is used for elaboration of the hardware lightning trigger. For this purpose the output signal comes, in parallel with ADC, to an amplitude discriminator which generates the trigger pulse each time when the detected amplitude of electromagnetic emission occurs above some predefined threshold. This trigger provokes recording of the next data series at both receivers of electromagnetic radiation, as well as at other detector systems it is connected to.

%VLF 100km-10km     3-30 kHz
%MF  1000m-100m   300-3000 kHz

The strength of the local electric field in thunderstorm time was measured by a ``field-mill'' kind sensor, similar to what is described in \cite{boltek}. This sensor is installed at the main territory of Tien Shan Mountain Station, $\sim$1.5\,km apart, and $\sim$400\,m below the high altitude detector point. The strength of detected field is encoded in analog form by the voltage of output signal of the field sensor; so that any level \textit{above} zero at this output means the presence of a positive electric charge in the spatial region just over the sensor, as well as possibility of electron acceleration in \textit{downward} direction between the main negative charge region in the middle of thundercloud and its lower positive charge distribution area, in the manner of as how it was described by \cite{thundermounts4,thundermounts7}.

At measurement time, the analog signal of the electric field sensor was digitized by a 12-bit ADC unit integrated into an STM32 microcontroller. The recording of these ADC data was realized similarly to the case of particle detectors: there were two parallel datasets, the monitoring data of the slow field variation with one second periodicity, and the fast time series of the field strength behaviour fixed with a 200\,$\mu$s resolution. The high resolution series were strictly bound to the lightning trigger generated by the discriminator of the electromagnetic emission signal.

The distance from the detector system to the region of electric discharge in recorded events can be roughly estimated by the time delay between the lightning flash and the arrival of thunder sound. For this purpose a microphone was installed at the high altitude point whose analog signal was digitized with another ADC unit, quite in the same way as what was applied for the electric sensor.

\section{Experimental results}

\subsection{The prolonged ground level enhancement event of radiation intensity}

\begin{figure*}
\centering
\includegraphics[width=0.49\textwidth, trim=8mm 0mm 0mm 0mm]{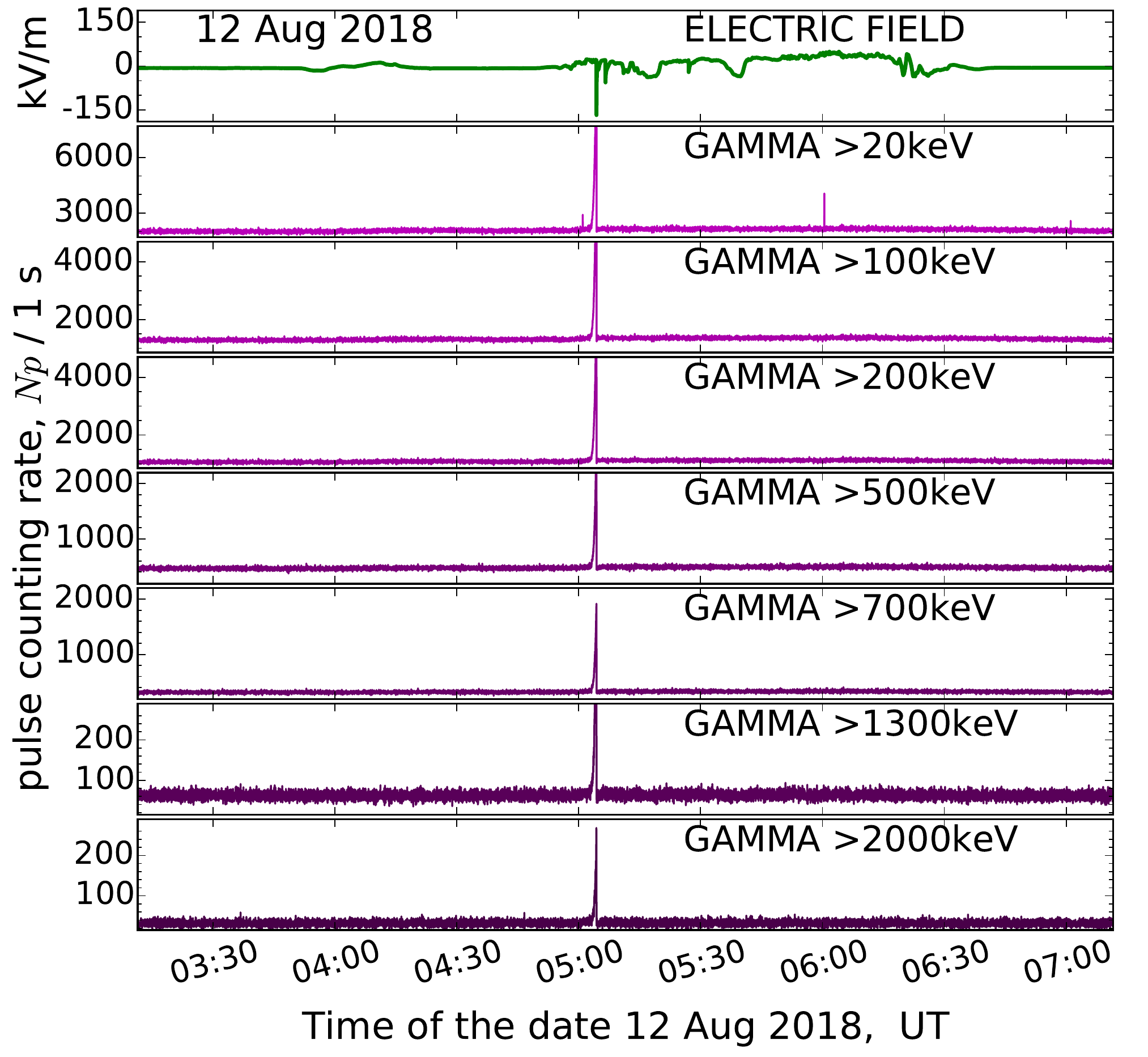}
\includegraphics[width=0.49\textwidth, trim=0mm 0mm 8mm 0mm]{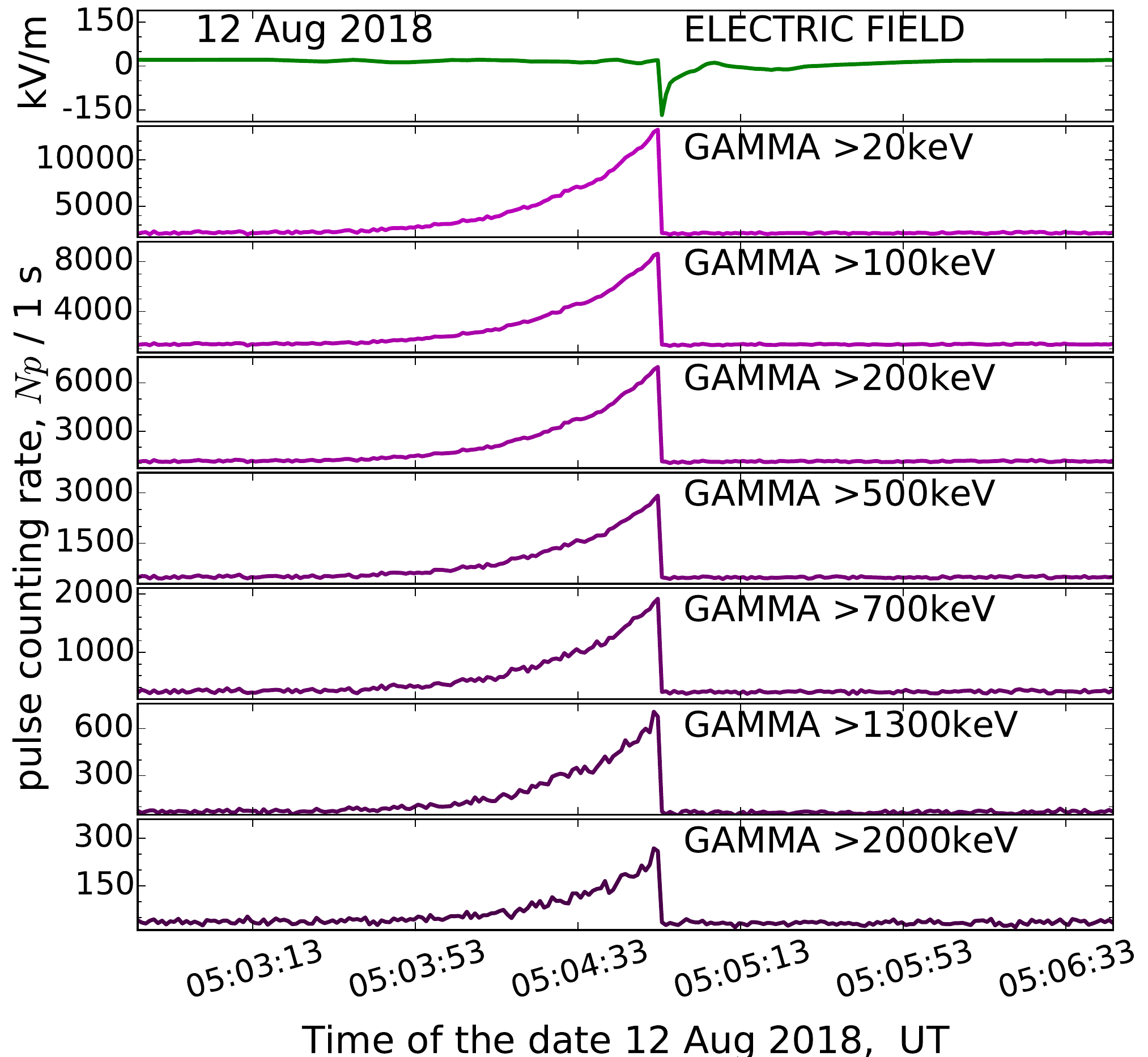}
\caption{The data of the gamma ray intensity monitoring made during the day of August~12,~2018. The timescale of the right frame is stretched around the moment of 05:05\,UT. Ordinate axes are graduated in the number of detector pulses $N_p$ obtained in a 1\,s long time interval.}
\label{figimonitogammo}
\end{figure*}

\begin{figure*}
\centering
\includegraphics[width=0.49\textwidth, trim=8mm 0mm 0mm 0mm]{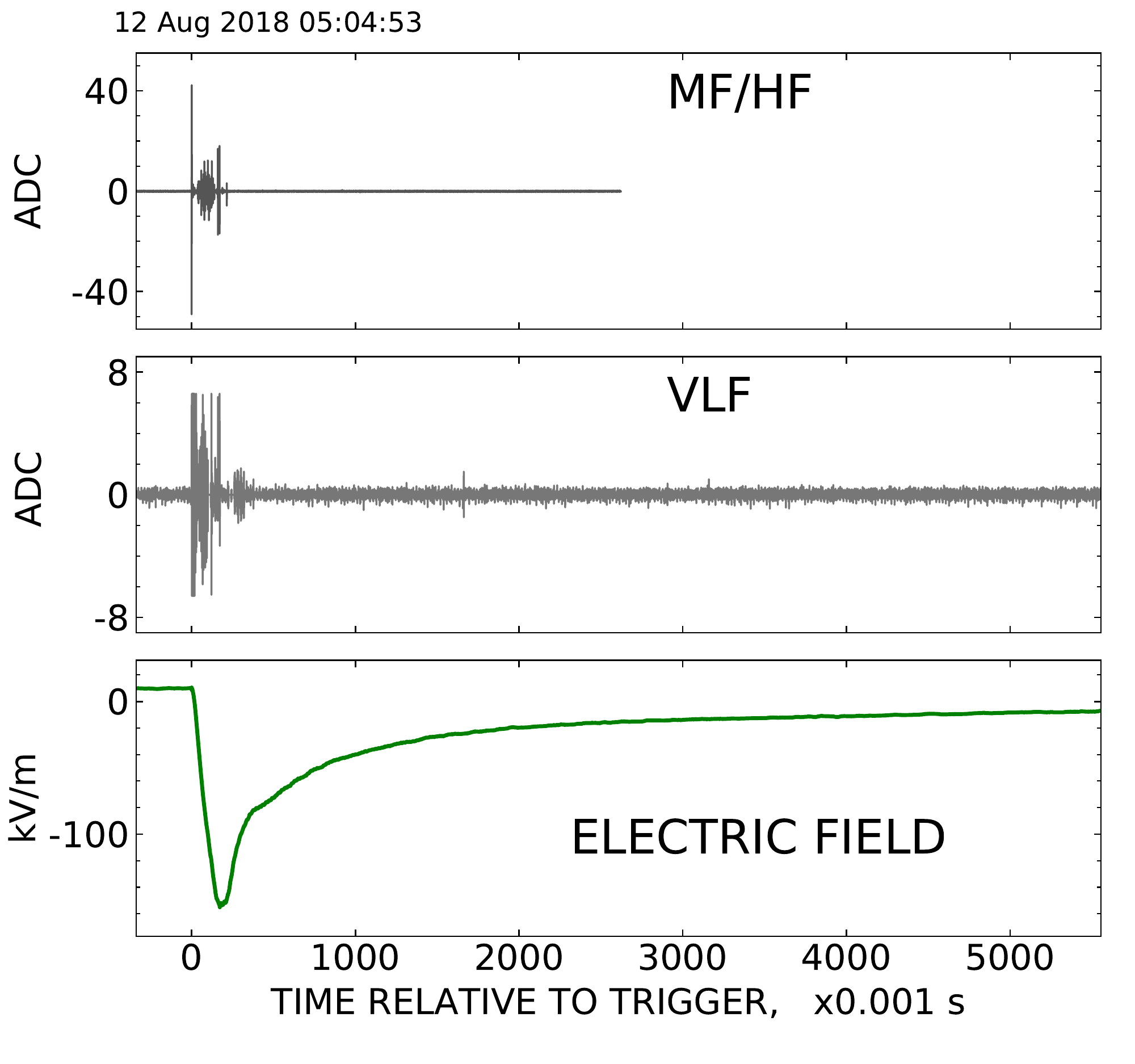}
\includegraphics[width=0.49\textwidth, trim=0mm 0mm 8mm 0mm]{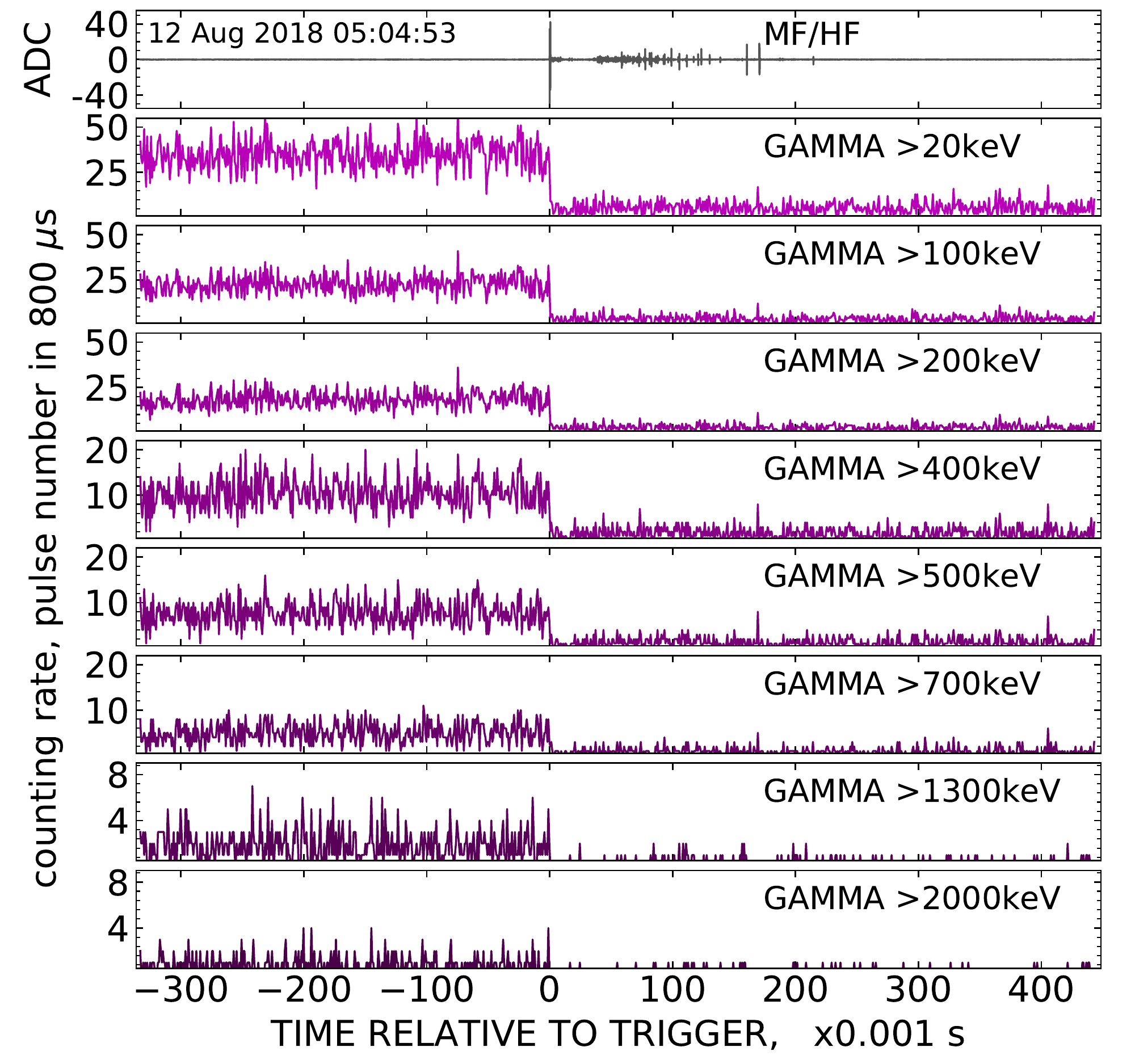}
\caption{Left: the local field variation and synchronous waveforms of the MF/HF and VLF electromagnetic emission (expressed in arbitrary ADC codes) after the moment of TGE termination. Right: high resolution records of the gamma radiation counting rates around the  lightning discharge which has terminated the TGE. Zero points of abscissa axes correspond to the moment of the lightning trigger caused by the discharge.}
\label{figimonitogammohighres}
\end{figure*}

\begin{figure*}
\centering
\includegraphics[width=0.5\textwidth, trim=4mm 0mm 4mm 0mm]{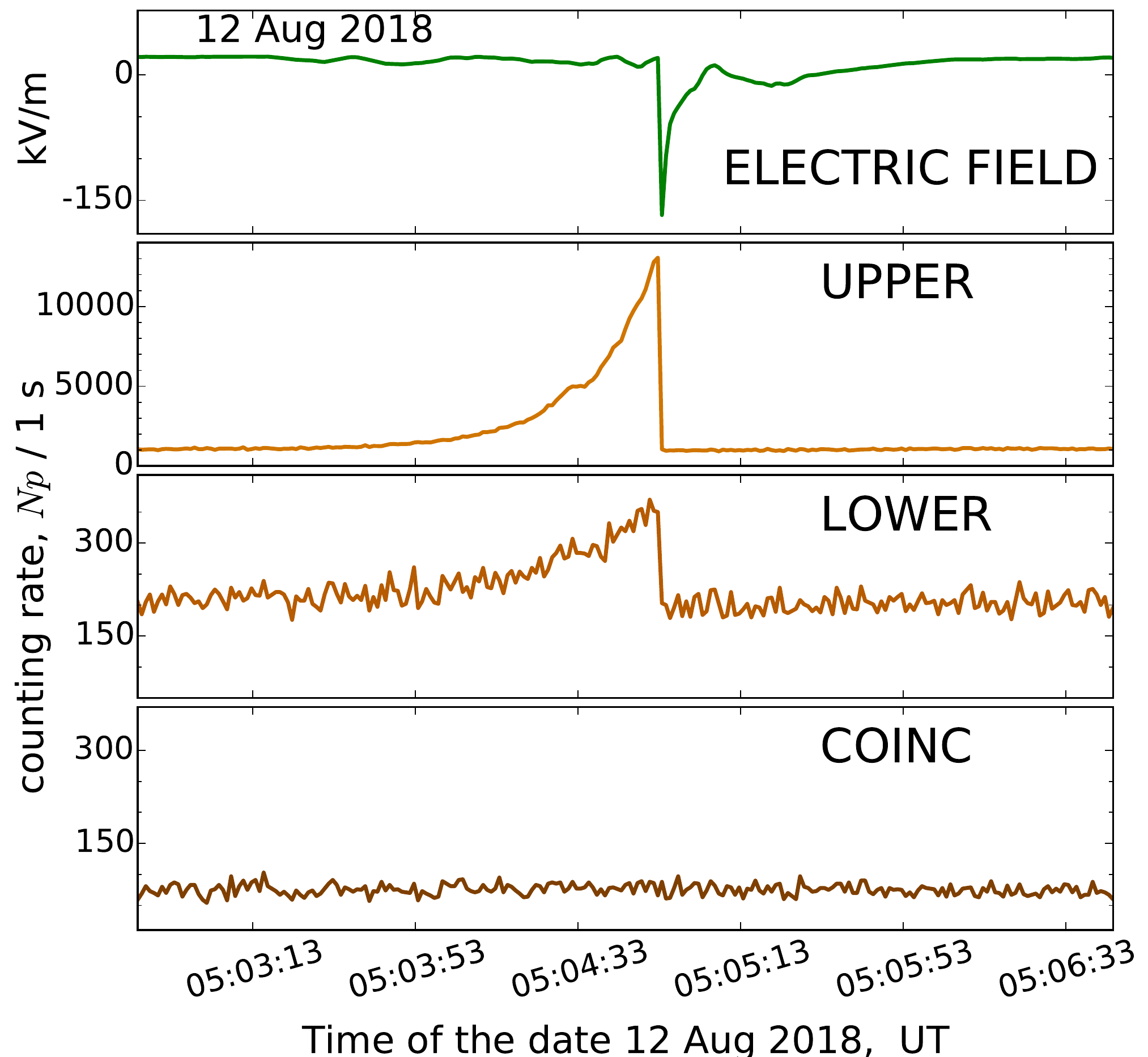}
\includegraphics[width=0.49\textwidth, trim=0mm 0mm 10mm 0mm]{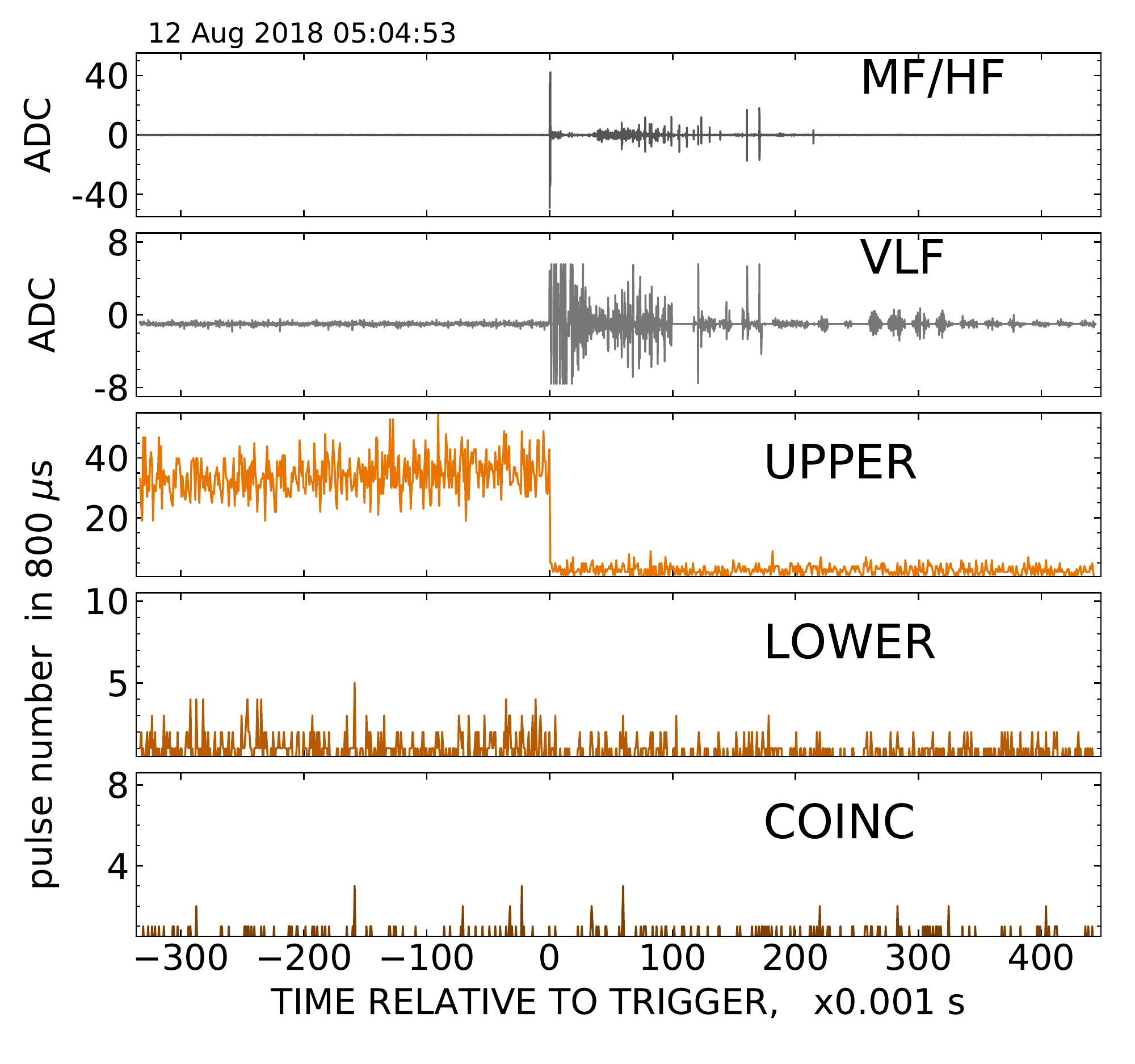}
\caption{The counting rates in the channels of the charge particles telescope (see the text), together with the records of the electric field and of the MF/HF and VLF electromagnetic emission at the time of TGE event. Left plot---the monitoring data with a 1\,s time resolution; right plot---the 800\,$\mu$s resolution series synchronized with the lightning discharge trigger (which coincides with zero point of the time axis).}
\label{figimonitoele}
\end{figure*}

During the measurements made in thunderstorm time of the date of 12\,Aug\-ust\,2018 it was detected an extremely large rise of gamma radiation which had preceded a close lightning discharge. The monitoring records of radiation intensity taken aro\-und this moment (August~12, 2018, 05:04--05:05\,UT) are presented in the plots of Figure~\ref{figimonitogammo}. Here, the gamma ray intensity is expressed in the units of the signal pulse numbers $N_p$ registered every second with various energy thresholds of the high altitude gamma detector. As it was mentioned in Introduction, in literature similar events were commonly designated as the thunderstorm ground en\-han\-ce\-ments---TGE. The distance from the detector point to discharge region estimated by the time delay of thunder sound relative to lightning flash in this event was of about (100--200)\,m.

As it follows from Figure~\ref{figimonitogammo}, at the time of TGE event the relative excess of radiation intensity in all energy ranges was 5--6~times above its background level, and the total duration of the TGE was of about 1.5\,min. The rise of radiation intensity has started significantly \textit{before} a close lightning discharge which has caused a large negative jump of the local electric field, and terminated just with beginning of the latter. The peak amplitude of the whole field variation at the end of TGE event was of about (100--150)\,kV/m.

In the high resolution data plots of the left panel in Figure~\ref{figimonitogammohighres} the time history of the lightning discharge which has terminated the TGE can be traced precisely by the waveforms of its MF/HF and VLF electromagnetic emission. The time series of these signals were detected with high temporal resolution and strictly synchronized by the lightning trigger with the moment of discharge beginning (see Instrumentation section above). The values of the electromagnetic signal amplitude in these plots are expressed in the arbitrary ADC code units.

In the electric field panels of Figures~\ref{figimonitogammo} and~\ref{figimonitogammohighres} it is clearly seen that during all the time of the TGE event the field measured by the electric sensor was oscillating at a moderate positive level of about $+$(10--30)\,kV/m, which agrees with acceleration of an electron flux in direction to the earth surface. Seemingly, such indication can be explained by rather prolonged existence of a positively charged region at the bottom of the thundercloud above the sensor. Thus, all the TGE time the sensor remained being screened from the influence of the main negative charge layer of the cloud, and the general field configuration was appropriate for acceleration of electrons in downward direction.

Just at beginning moment of the terminating lightning discharge the polarity of detected field had quickly changed to opposite (i.e. negative) and fell down to the level of $-$(120--130) kV/m. This jump agrees again with sudden dissipation of the lower positive charge caused by lightning current, and with subsequent response to the main negatively charged layer of thundercloud from the side of the field sensor.
Total duration of the TGE terminating discharge was of about (0.3--0.5)\,s only, but the complete relaxation of the electric field after the jump to its initial level in vicinity to zero has taken an order of magnitude longer time.

The high resolution records of the intensity of gamma radiation signal measured just before and after the terminating discharge are illustrated by the right panel plots of Figure~\ref{figimonitogammohighres}. It is seen there a rather abrupt disappearance of any gamma radiation signal just at the moment of final lightning. More on this subject follows below.

\begin{figure*}
\centering
\includegraphics[width=0.8\textwidth, trim=0mm 0mm 0mm 0mm]{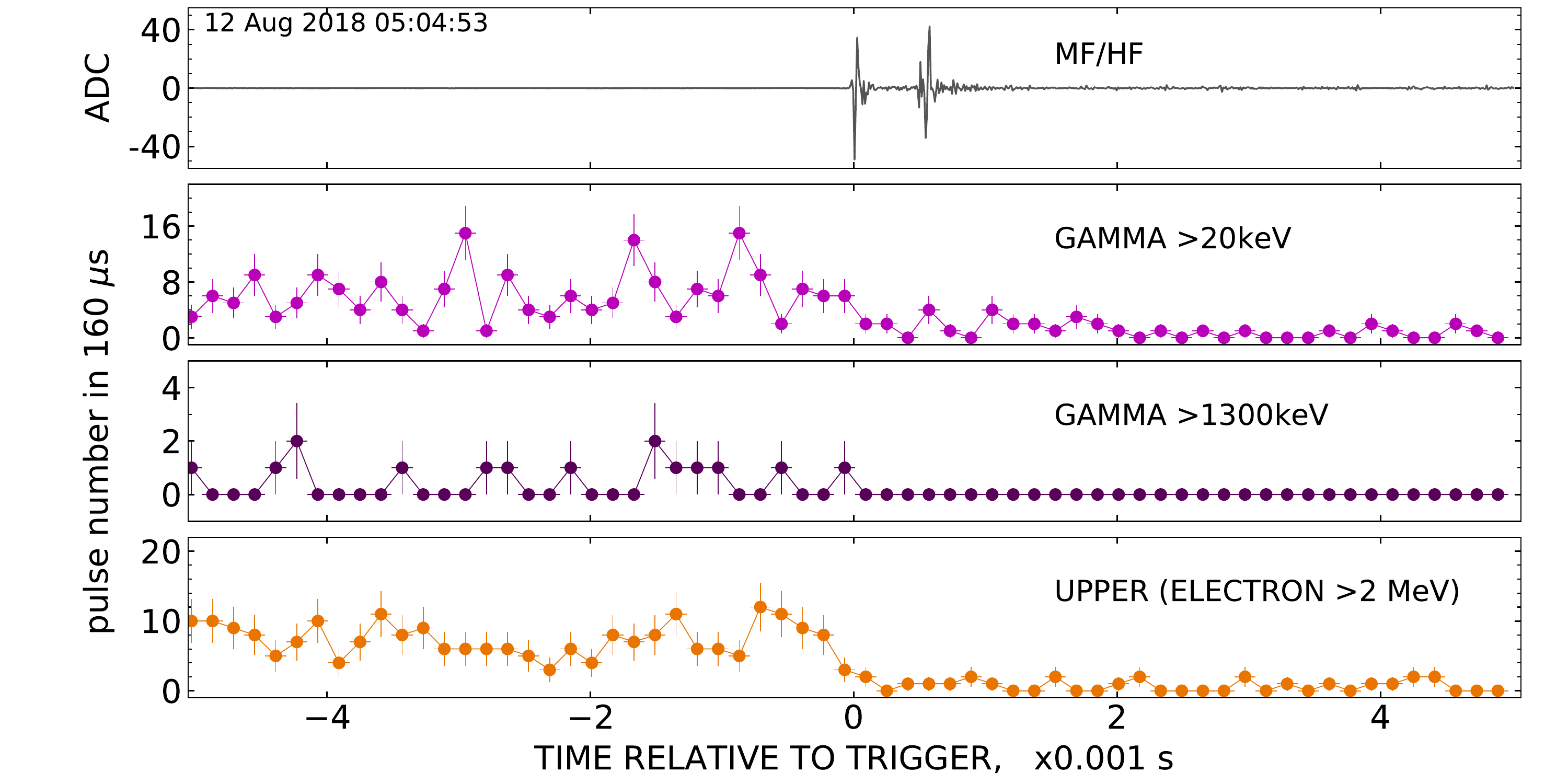}
\caption{The radiation counting rate records zoomed around the TGE termination moment. Time resolution is 160\,$\mu$s. Vertical bars correspond to statistical error of the intensity measurement.}
\label{figimonitogammozoomed}
\end{figure*}

Time series of the charged particles signal registered by the high altitude telescope detector in the time of the TGE event
are presented in Figure~\ref{figimonitoele}. As it was explained above, the signals of the upper scintillator layer of this telescope correspond to relativistic electrons with the energy of a few MeV order. According to the \textit{UPPER} designated panels of Figure~\ref{figimonitoele}, the peak amplitude of the intensity increase of such particles during the TGE is of $\sim$(10--12)~times as much as its usual background. The rise of the charged particles intensity terminates just before the moment of the final lightning discharge, simultaneously with TGE gamma radiation.

In principle, the signal from the electron component of the thunderstorm activity caused radiations could be imitated by the charged products of gamma ray interaction, such as Compton scattering and the like. Relative admixture caused by this effect into sum signal of our plastic scintillator detector can be evaluated by comparison of the \textit{UPPER} an \textit{LOWER} labelled panels in Figure~\ref{figimonitoele}: while the former demonstrates an order of magnitude high rise at the maximum of TGE event, the peak amplitude of the latter above its background was of about (50--100)\% only. Because of a thick absorber layer between these two scintillators, the lower one should be sensitive mostly, indeed, to products of gamma ray interaction both within the absorber and in outer environment (the soil beneath, \textit{etc}), while any excess over its signal detected by the upper scintillator corresponds to the pure deposit from the side of charged particles in thundercloud coming from above.

Most significant is the null response in \textit{COINC} panels of Figure~\ref{figimonitoele}. Since, as it was pointed out in Instrumentation section, the probability of gamma radiation detection by a single plastic scintillator is essentially below 1\%, it remains only negligible probability for any particular gamma ray quantum to interact twice in the upper and lower detectors of the telescope set-up with producing of charged particles and generation of scintillation pulses in both layers within the same gate time of (2--3)$\mu$s. Hence, the detection of coincidence pulses between these two layers, if any, would be a passage sign of a high-energy, $\gtrsim$(80--100)\,MeV, charged particle, and the lack of such signals in Figure~\ref{figimonitoele} means the absence of energetic electrons acceleration in the time of considered TGE event.

The sharpness of the radiation decay at termination moment of the TGE event is illustrated once more by Figure~\ref{figimonitogammozoomed}. As it follows from the plots presented here, any surplus emission above the usual background, both for the gamma ray and electron components, had extincted very quickly just at the beginning of the final discharge. Evidently, the characteristic time of this disappearance at any rate was below 160\,$\mu$s length of a single interval of intensity measurement.

\subsection{The short-time radiation burst from a lightning discharge}

\begin{figure*}
\centering
\includegraphics[width=0.49\textwidth, trim=0mm -25mm 7mm 0mm]{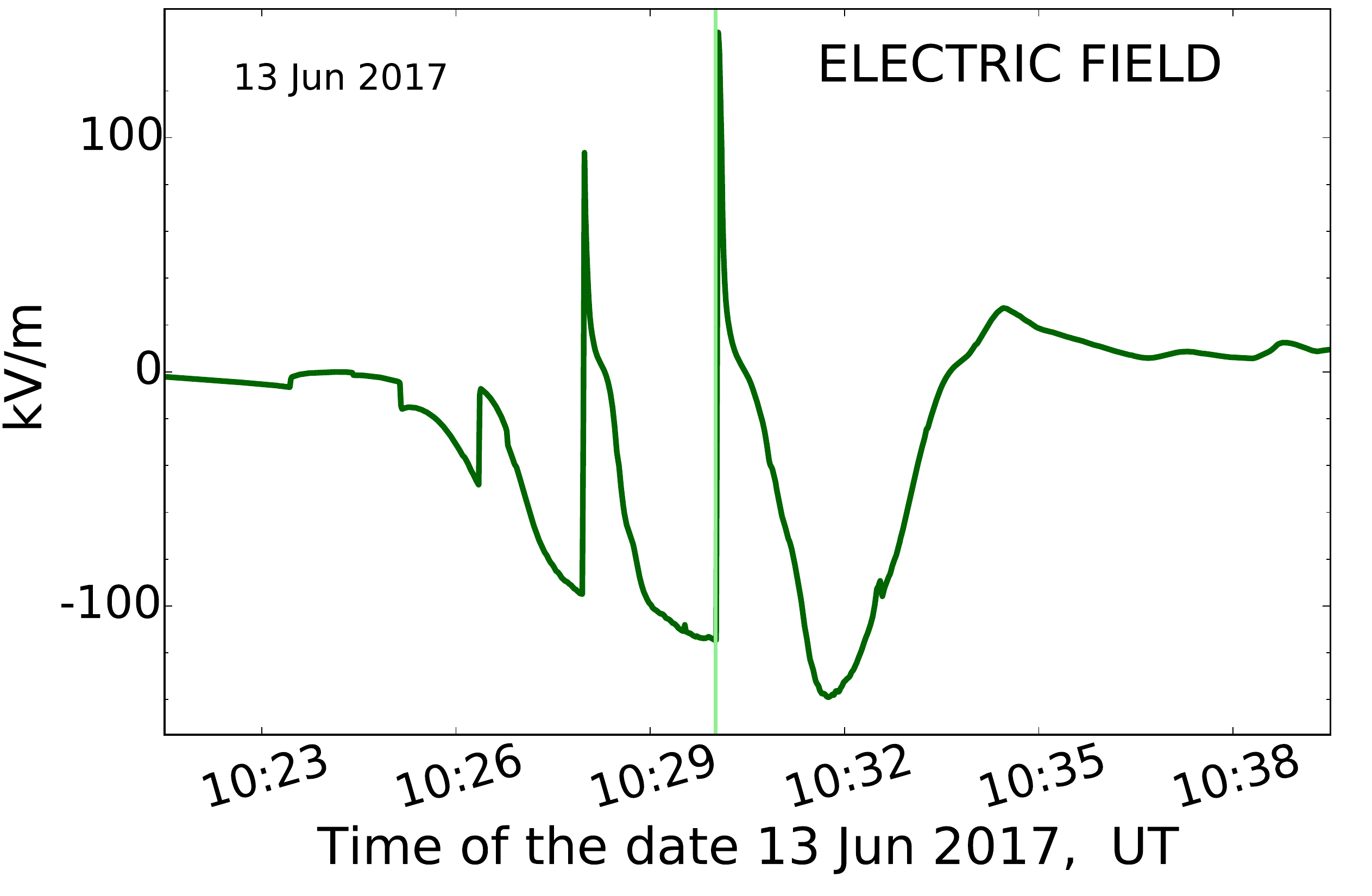}
\includegraphics[width=0.49\textwidth, trim=0mm 0mm 7mm 0mm]{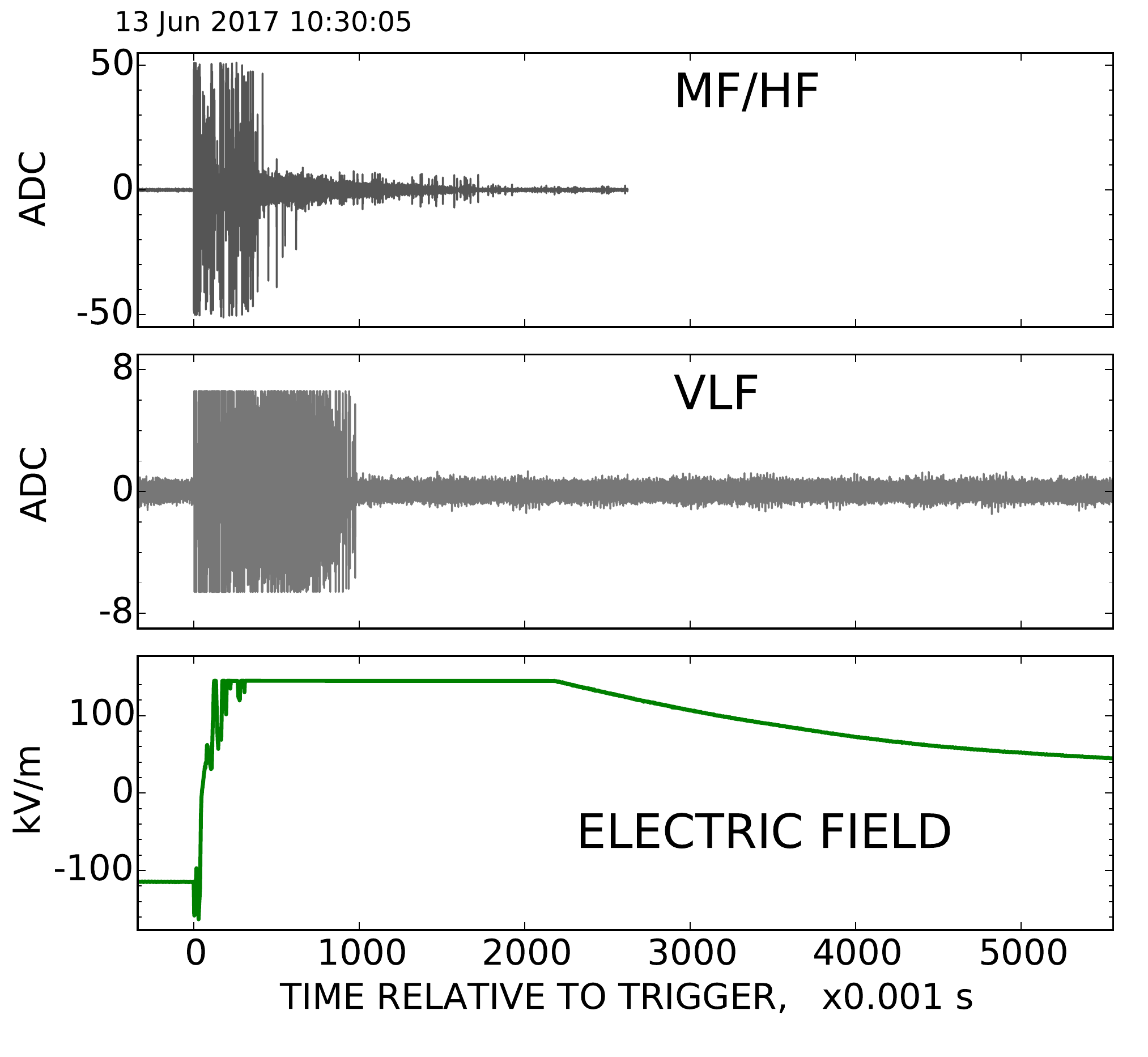}
\caption{Left: the behaviour of local electric field around the time of the June 13, 2017, 10:30:05\,UT radiation burst event; the moment of the burst is marked with a vertical line. Right: the variation of electric field and the records of electromagnetic radiation from the atmospheric discharge. Zero point of abscissa axis in the right panel corresponds to the moment of discharge initiation notified by a lightning trigger signal.}
\label{figitgfele}
\end{figure*}

\begin{figure*}
\centering
\includegraphics[width=0.49\textwidth, trim=10mm 0mm 0mm 0mm]{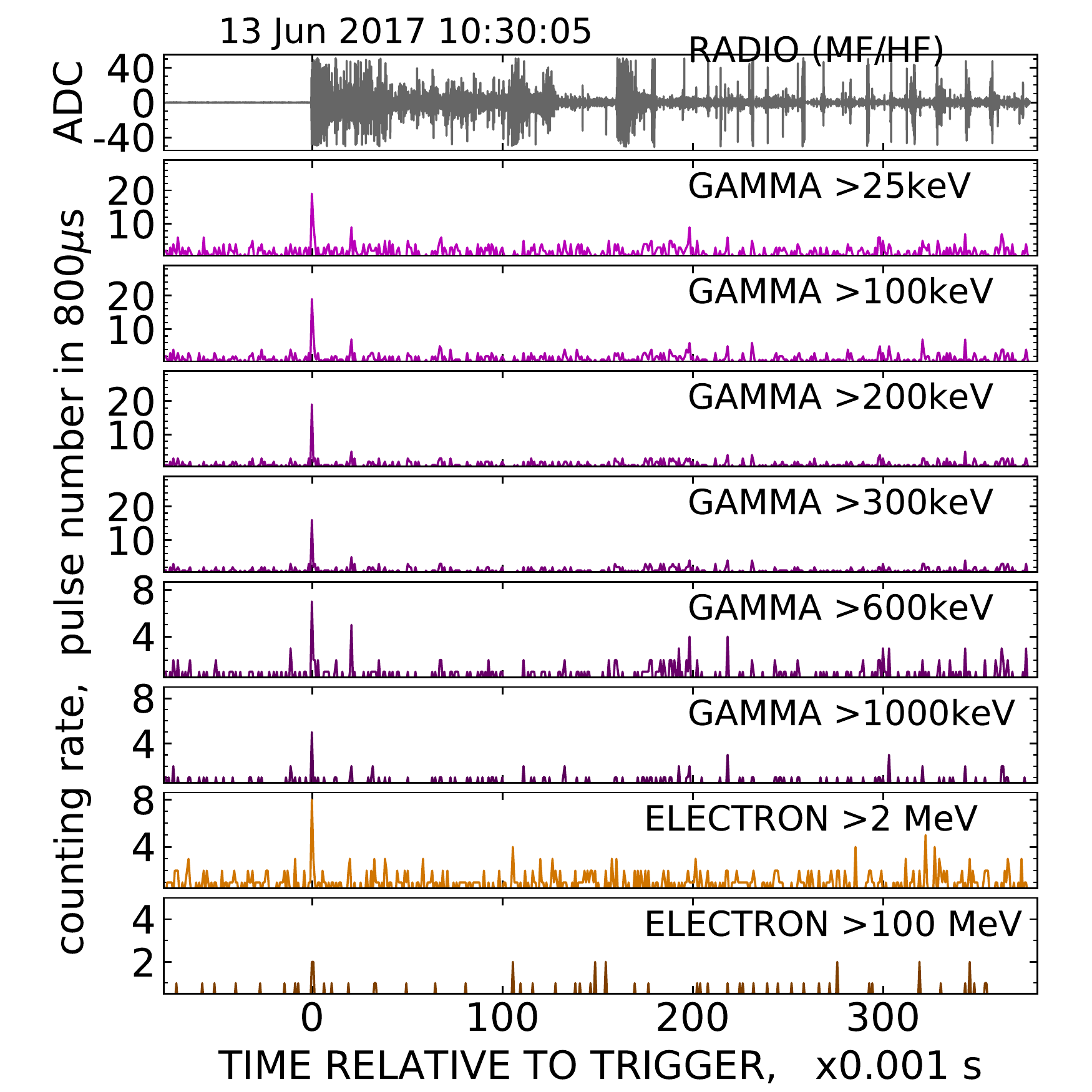}
\includegraphics[width=0.49\textwidth, trim=0mm 0mm 10mm 0mm]{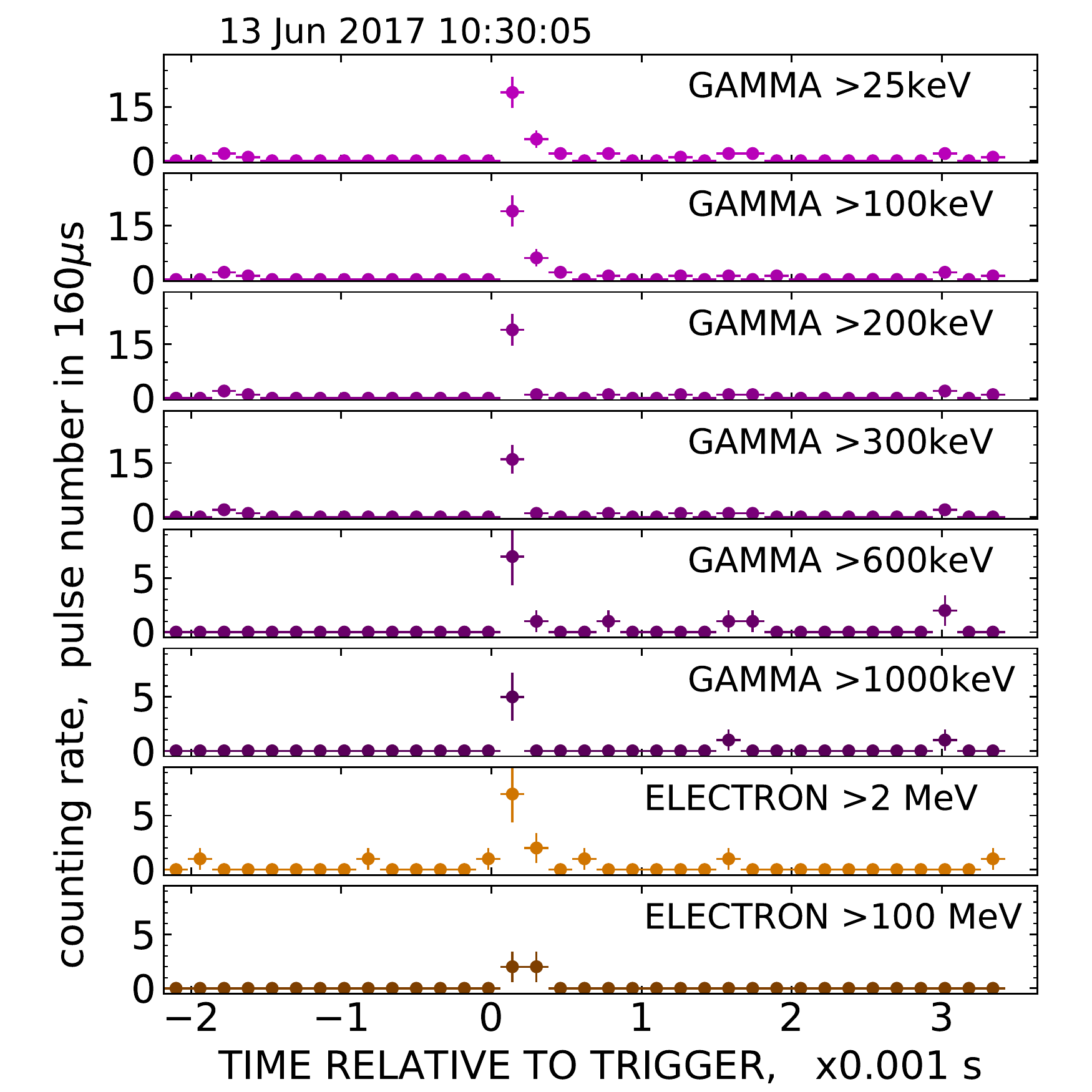}
\caption{High resolution time series of the  gamma ray and charged particles counting rates at the time of a close radiation burst. Time axis in the right frame is stretched around the moment of the lightning trigger. Time granularity of signal intensity measurements is 0.8\,ms in the left plot, and 0.16\,ms in the right, zero point of the time axes corresponds to initiation moment of the lightning.}
\label{figitgf}
\end{figure*}

Another kind of thunderstorm activity events detected at Tien Shan Mountain Station is presented by a short time outburst of the intensity of gamma rays and accelerated electrons which occurred simultaneously with a close lightning discharge. As it is shown in Figure~\ref{figitgfele}, at the moment of June 13, 2017, 10:30:05\,UT a sudden jump of field tension between the limits from $\mathcal{E}\approx -130$\,kV/m to $\mathcal{E}\gtrsim +100$\,kV/m was accompanied by intensive electromagnetic emission from an atmospheric discharge which has started just at beginning of the field variation. According to the delay of thunder sound, the discharge distance from the high altitude detector point in this event was below~100\,m.

In contrast to the TGE event considered above, from the high resolution time series of the gamma ray counting rates presented in Figure~\ref{figitgf} it follows that the intensive radiation flash in present case just \textit{coincided} with the initiation moment of the electric discharge. The whole duration of the radiation increase was of a sub-millisecond order or shorter: as it is seen in the right panel plot of Figure~\ref{figitgf}, most part of the flash goes in a single 160\,$\mu$s long interval of the intensity measurements, and it does terminate completely up to the end of the next interval.

Together with gamma-radiation, synchronous outbursts both of low-energy ($\geqslant$2\,MeV) and high-energy ($\gtrsim$100\,MeV) electrons are seen in the plots of Figure~\ref{figitgf}. These data correspond to the \textit{UPPER} and \textit{COINC} channels of the charged particles telescope detector.

Relative amplitude of the excessive radiation peak calculated as \mbox{$(I_{peak}-I_{bckgr})/I_{bckgr}$} over the data of Figure~\ref{figitgf} (where $I_{bckgr}$ is the background counting rate) varies in the limits of (1500--2000)\% for gamma rays, 1000\% for $\geqslant$2\,MeV electrons, and up to 4000\% for the high energy electron deposit.
%	mean	peak	rel
% 1*	1742 *	27	18
% 2	1595
% 3	1351
% 4	1180
% 5*	1143 *	20	21
% 6*	952 *	17	21
% 7	791
% 8*	596 *	8	16
% 9	517
% 10*	469 *	6	15
% 11	350
% 12*	303 *	5	20
%
% u	940	9	11
% c	122	4	40
% >>> def rel( a, b ) :
% ...   return ( a / (800 * 1.e-6) - b ) / b ;
% ...
%>>> rel( 27, 1742 )
% 18.374282433983925

Therefore, the phenomenological features of the radiation burst  detected at initiation stage of the June 13, 2017, 10:30:05\,UT lightning discharge do cardinally differ from the properties of the former TGE event. On account of an order of magnitude higher relative amplitude, the presence of high energy electrons, and its generally short-term time profile, the observed radiation excess can be attributed to a quite another class of the atmospheric electricity phenomena. Evidently, this is a representative of  short time intensive bursts of hard radiation coinciding with lightning. As it was discussed in Introduction, similar events were multiple times reported by various experimental groups, and presently it is supposed that such transient bursts originate from stepped leaders which arise in the course of lightning development. As it was pointed out in \cite{dwyer_review2012} and in literature cited therein, in the basis of these phenomena it may be the ``cold'' runaway breakdown reactions which take place in vicinity to the tops of the electric discharge streamers in developing lightning leader.

With application of streamer model the origin of the burst event detected on June~13, 2017 could be ascribed to particles avalanches which were originating through the ``cold'' runaway process just at initiation stage of discharge, and then were moving in downward direction and growing further in a strong but short living field of appropriate direction. A source of such field might be sudden diminishing of the total charge of main negative layer in the middle of thundercloud which caused an appearance of a transient positively charged region in its lower part. In the plots of Figure~\ref{figitgfele} this field leaved its traces as a narrow peak of positive polarity around the moment of 10:30:05\,UT. It can be seen there that the width of the peak was of a few seconds only, but this is much longer then the whole duration of the lightning, not to mention the hard radiation burst. After fast dissipation of the lower positive region the field registered locally in the earth vicinity returned back to its initially large negative level which seemingly mirrored the main negative charge in the cloud above.

It is probable that even in presence of a favorable field structure at discharge moment successful observation of intensive gamma ray burst demands fulfillment of rather specific conditions, since, as it can be traced in Figure~\ref{figitgfele}, during the same thunderstorm, and even within the same thundercloud system there were detected at least two similar jumps of the local electric field, those at 10:26:25\,UT and 10:28:00\,UT, and the amplitude of the latter was quite comparable with that of the final 10:30:05\,UT radiation event. In the moments of both preceding field jumps it was detected, indeed, an intensive electromagnetic emission from the nearby lightnings similar to the waveforms in the \textit{MF/HF} and \textit{VLF} labelled panels of Figure~\ref{figitgfele}, but none of them was accompanied by any detectable flash of hard radiation.
% to varify in can be done the next:
%shethunder2013=> SELECT datep,timeut from thunderelepupscans where datep='13.6.2017' order by timeut ;
%shethunder2013=> SELECT datep,timeut from thunderradiscans where datep='13.6.2017' order by timeut ;
% the '13.06.2017 | 10:28:00' event records are present indeed.
The rarity of the burst observations even at close lightnings may arise because of narrow angular distribution and fast absorption of the high energy emissions from electron-photon avalanche developing in thunderclouds.

It should be stressed that because of a rather prolonged dead time of the applied detectors which was of the order of a few microseconds, in the ``trigger'' (zero-point) time interval of Figure~\ref{figitgf} we have to deal most probably with an overlap of signals from many elementary streamers which were developing successively in lightning leader, so all conclusions made below on energy spectra, intensity of particles fluxes, \textit{etc} relate mostly to average streamer characteristics.

Some more considerations on possible origin of the June 13, 2017 event follow in the end of next section~\ref{sectifluxi}.

\section{Discussion}

\subsection{Energy spectra and radiation fluxes}
\label{sectifluxi}

\begin{figure*}
\centering
\includegraphics[width=0.49\textwidth, trim=10mm 0mm 0mm 0mm]{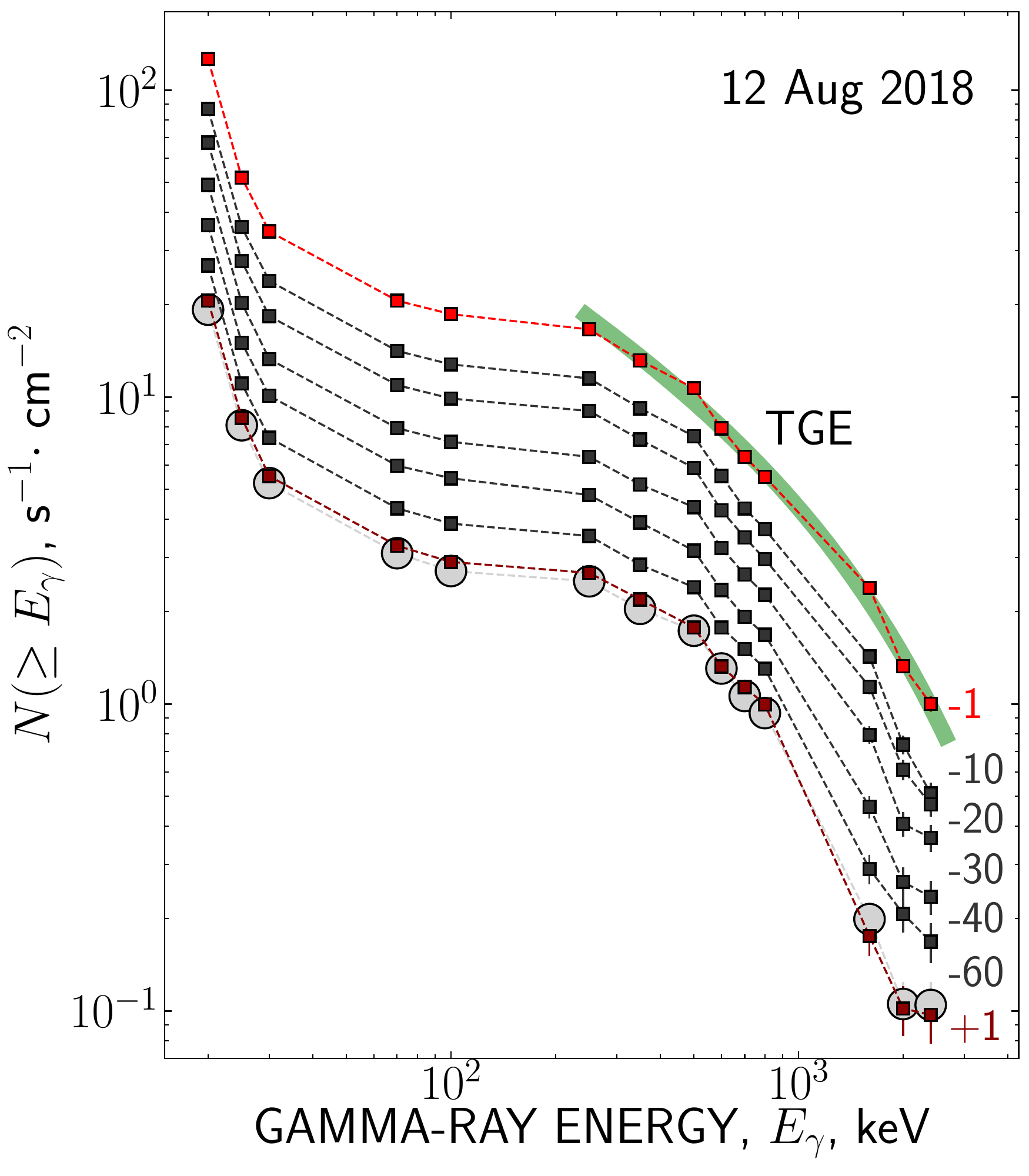}
\includegraphics[width=0.49\textwidth, trim=0mm 0mm 10mm 0mm]{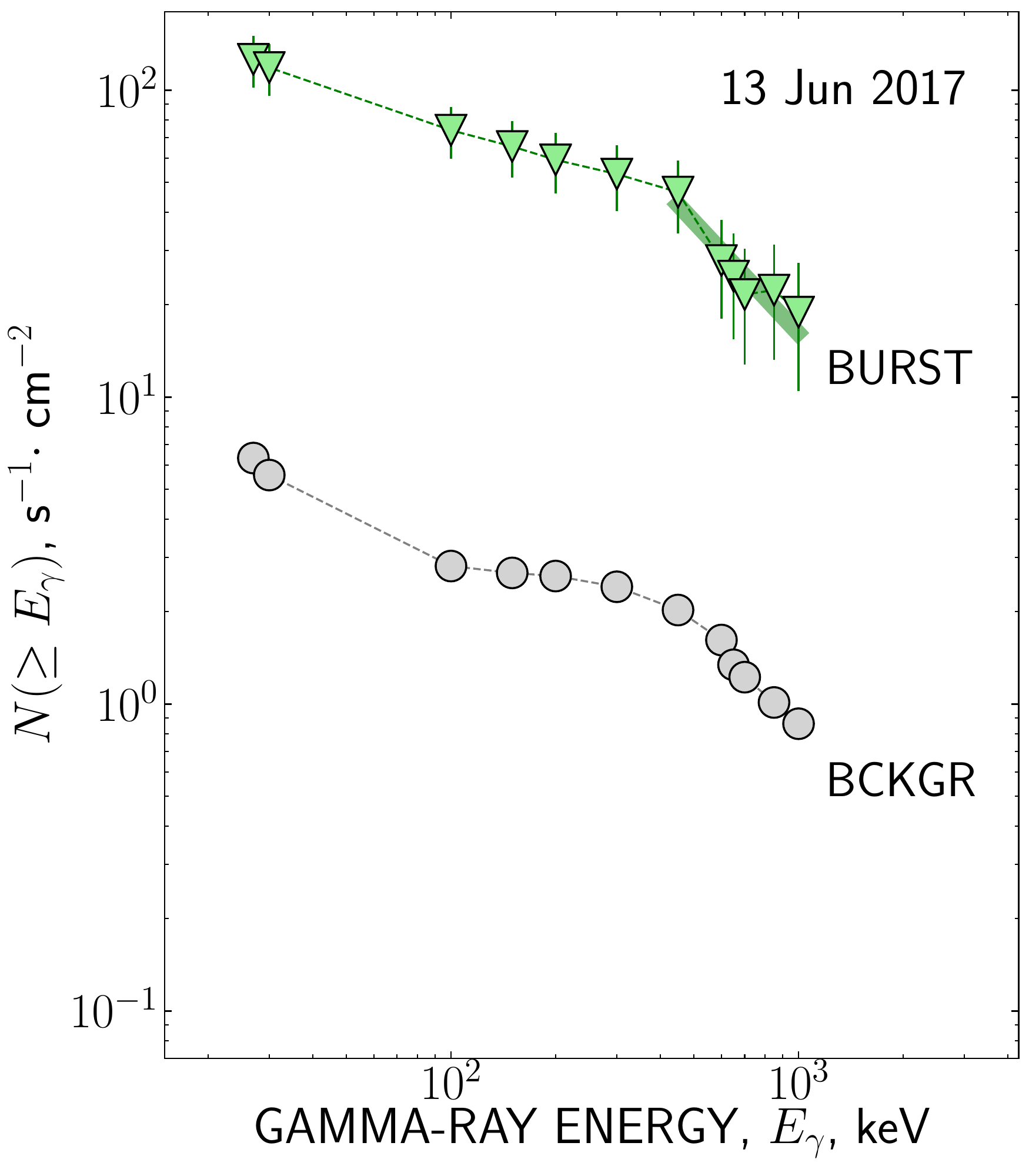}
\caption{Left: integral spectra of TGE radiation. The numbers beside curves mean the time delay (in seconds) relative to the final discharge moment. Right: the spectrum of the short radiation burst (triangles). Background spectra are marked with circles. The smooth continuous lines correspond to analytic approximation of the experimental data points (see text).}
\label{figigammospc}
\end{figure*}

The monitoring type data presented in Figure~\ref{figimonitogammo} permit to calculate the momentary energy spectra of detected gamma rays, each of which corresponds to some particular period of time preceding the moment of TGE termination. A set of such partial distributions gives an opportunity to trace any possible evolution of the integral spectrum $N(\geqslant E_\gamma)$ of gamma radiation in the course of the TGE event development.

A number of such energy spectra is presented in the left frame plot of Figure~\ref{figigammospc}. By their calculation it was taken into account the total sensitive area of scintillator crystal used in the gamma detector (570\,cm$^2$), as well as the distribution of gamma ray detection efficiency for different $E_\gamma$ from  Figure~\ref{figigammoeffici}.

The amplitude of the leading points in all spectra of Figure~\ref{figigammoeffici} can be somewhat underestimated because of attenuation of the soft gamma rays within the 1\,mm thick iron walls of the shielding box which surrounds the gamma detector from outside; nevertheless, as it can be calculated by subtracting the background intensity from the intensity of corresponding spectra points in the left panel plot of this figure, just before the termination moment of the TGE an absolute increase of the low-energy radiation flux \mbox{$E_\gamma\gtrsim 30$\,keV} was of about (30--35)\,s$^{-1}$cm$^{-2}$ order. In the energy range of (100--300)\,keV the peak intensity of the TGE connected radiation excess occurs of \mbox{(15--18)\,s$^{-1}$cm$^{-2}$}, and  around~1\,MeV it was of \mbox{$\sim$(5--7)\,s$^{-1}$cm$^{-2}$} only.

As it is seen in the left panel plot of Figure~\ref{figigammospc}, generally the gamma ray spectrum had retained untouched its shape during the whole TGE event, and the only observable change of its form was a gradual rise of the absolute radiation intensity. Just after the final lightning discharge this intensity falls down again to the level of the background spectrum (which is shown by circles in Figure~\ref{figigammospc}). Similar behaviour of the gamma ray spectra detected during TGE times was mentioned in \cite{thunderaragats2019}.

The TGE radiation spectra in Figure~\ref{figigammospc} have a rather complicated form but in the range above a few hundreds of keV they can be approximated by a function \mbox{$f(E_\gamma)\sim E_\gamma^{-\alpha}\cdot \exp(-E_\gamma/\varepsilon))$}; analogously to as how it was done by \cite{thundertgejapan2019} for the gamma rays originating from an electron-photon avalanche, and with account to the Compton scattering process. In the present case the best fit parameters are \mbox{$\alpha\approx 0.68$} and \mbox{$\varepsilon\approx 1.6$\,MeV}, which approximation is shown by a smooth continuous line in Figure~\ref{figigammospc}. The resulting estimation of $\alpha$ falls well into the range of power indices reported for the TGE events which were observed by Aragats group, \textit{e.\,g.} in  \cite{thunderaragats2014_on_mos}. As it is explained in that publication, such power spectra may be accounted for by the MOS type process of particles interaction in thunderclouds with a comparatively small electric field which remains below the critical energy threshold of the large-scale RREA discharge. To the same conclusion leads as well the value of $\varepsilon$ parameter above which occurs much below the mean energy of RREA particles (\mbox{$\simeq$7\,MeV}). This differs the present TGE event from the other analogous cases, such as reported in \cite{thundermounts1,tsuchiya2007,aragatstge2011b}, where it was claimed existing of a tail of hard radiation which lasted up to tens of MeV.

At the same time the presently detected spectra can not exclude the possibility of electrons acceleration due to the ``cold'' runaway process. As it was noted \textit{e.\,g.} by  \cite{dwyer-arabshhi_2015}, in that case there does not exist any typical average energy of resulting gamma emission because of strong dependency of avalanche development on particulars of specific configuration of the local electric field in thundercloud. Nevertheless, it seems very unlikely that the field with any strength necessary for initiation of the ``cold'' discharge could sustain some noticeable time in the air, so the MOS model remains to be considered as an only probable mechanism for explanation of our prolonged radiation event.

The energy spectrum of the gamma ray flash detected at the moment of a close lightning discharge on June~13, 2017 is presented in the right panel plot of Figure~\ref{figigammospc}. This spectrum was calculated according to the summed pulse count values in three consecutive 160\,$\mu$s long intervals, the first of which coincides with zero point of abscissa axis in the high resolution data series in Figure~\ref{figitgf} (\textit{i.\,e.} only those signals participate in calculation which have come during a $\simeq$500\,$\mu$s long time period just after the lightning trigger). Unfortunately, at that time  in the DAQ system of the gamma detector there were not anticipated any data channels with registration threshold high enough to analyze the signal from hard gamma rays, so the measured spectrum terminates at the point of 1\,MeV. Nevertheless, it is seen in Figure~\ref{figitgf} that at the energies of \mbox{$E_\gamma \gtrsim (100-300)$\,keV} the absolute amplitude of the burst spectrum is 2--3 times above the peak radiation intensity which has been detected at the time of the TGE event, so the absolute deposit of the burst radiation is of $\sim$50\,s$^{-1}$cm$^{-2}$ in the range of (100--300)\,keV, and of $\sim$20\,s$^{-1}$cm$^{-2}$ at~1\,MeV. For soft gamma rays with $E_\gamma\gtrsim 30$\,keV an absolute excess of the burst radiation above the background occurs of the order of $\sim$100\,s$^{-1}$cm$^{-2}$.

Besides, the spectrum of the short-time burst event seems to be somewhat harder then that of the TGE: above a few hundreds of keV it generally corresponds to a power approximation of the $\sim$$E_\gamma^{-1.2}$ type (which is shown by a continuous sloped line in the right frame plot of Figure~\ref{figigammospc}), instead of a fast exponential fall down in the case of TGE. A similar power shape of the integral radiation spectra, and with close value of its slope index has been just detected in the energy range of (0.1--10)\,MeV in former measurements at Tien Shan station, such as those presented in \cite{thunderour2016}. Differential energy spectra of a power shape $dI/dE_\gamma\sim E_\gamma^{-\alpha}$ with $\alpha$ within the limits of 2.0--2.5 were reported also among the results of other experimental groups, \textit{e.\,g.} by \cite{thundermounts7,thundermounts3,thundermounts4}, for a number of gamma ray flashes detected at the times of lightning discharges. These data are also compatible with our present result.

%
% ++++ 1.8.2020.
% electron backgrounds:    electron peaks:       electron absolute
%        upper   coinc      upper   coinc.        upper    coinc
% TGE     970/s   70/s      13000     -           12000      -
% burst   950/s  120/s  28000/19000 12000/8000  27000/18000 12000/8000

% `tge-upper` follows immediately from the monitoring data plot.
%  for the `burst` initial data are as the following:
%2667	90.56	19	19	19	16	7	5	7	2
%2668	90.72	6	6	1	1	1	.	2	2
%                                                             ------------
%                                                               9       4
%  9/320e-6 = 28000     4/320e-6 = 12000
%  9/480e-6 = 19000     4/320e-6 = 8000

Absolute deposit on the part of thunderstorm activity into the electron component of detected signals can be estimated immediately by the monitoring type records from Figure~\ref{figimonitoele} for the TGE event, and as a sum of pulse counts in corresponding time series of Figure~\ref{figitgf} for the short time radiation burst. (In latter case the numbers of the charged particles detector pulses detected in three 160\,$\mu$s intervals just after the lightning trigger were included into this summing, quite in the same manner as how it was done above by calculation of the gamma burst spectrum). In both cases the corresponding background count rates were subtracted from the intensity of thunderstorm connected signals, and the remaining differences were normalized to the 1\,m$^2$ sensitive area of the charge particles detector, and to duration of the pulse measurements period (\textit{i.\,e.} to~480\,$\mu$s in the case of the short radiation burst). The results of this procedure are of about $F_e\simeq 2$\,s$^{-1}$cm$^{-2}$ and $F_e \simeq 1$\,s$^{-1}$cm$^{-2}$ for the electron fluxes which were detected at the moment of the short radiation burst with energy thresholds, correspondingly, of $E_e\gtrsim 2$\,MeV and $E_e\gtrsim 100$\,MeV. As for the TGE event, the peak flux of low-energy electrons detected during the last second just before its termination was at a comparable level of 1.3\,s$^{-1}$cm$^{-2}$.

%>>> '%e' % ( 2.*10000 * (100**2) * (3*10**2) ) = '6.e+10'
%>>> '%e' % ( 2.*10000 * (100**2) * (3*30**2) ) = '5.e+11'
%>>> '%e' % ( 2.*10000 * (100**2) * (3*100**2) ) = '6.e+12'
%               ^-- from cm^{-2} to m^{-2}
%                           ^-- 100m distance to emission point to square
%                                       ^-- total outer area
%                                           of a 30/100m sized region
The electron flux estimate for the June~13, 2017 burst event permits to make an additional conclusion on its possible origin. Based on the above $F_e$ values, accepting $\sim$100\,m order distance to the source of detected emission, and ignoring attenuation of the electron flux on its way to detector one can deduce a lower possible limit for the total electrons number as $N_e\sim (10^{11}-10^{12})$, in supposition of the outer size of emitting region of about $(30-100)$\,m correspondingly. Such scale of $N_e$ is in reasonable agreement with typical multiplicities of electron production by stepped leaders which were reported by many experimental groups and can be found \textit{e.\,g.} in the review of \cite{dwyer_review2012}. This result confirms once more our initial assumption on a stepped leader process in close lightning as a source of  observed hard radiation in this case.
%babich_cascadesimu_2013
%tsuchia_2011_tgf_spectrum_from_dwyer

\subsection{An imprint of positron production}

%TODO
% burst event of 13.6.2017: differential spectra evolution during the whole scan to see if any dissipation of the 512keV line does take place over a second long time lapse?

\begin{figure*}
\centering
\includegraphics[width=0.49\textwidth, trim=0mm 0mm 0mm 0mm]{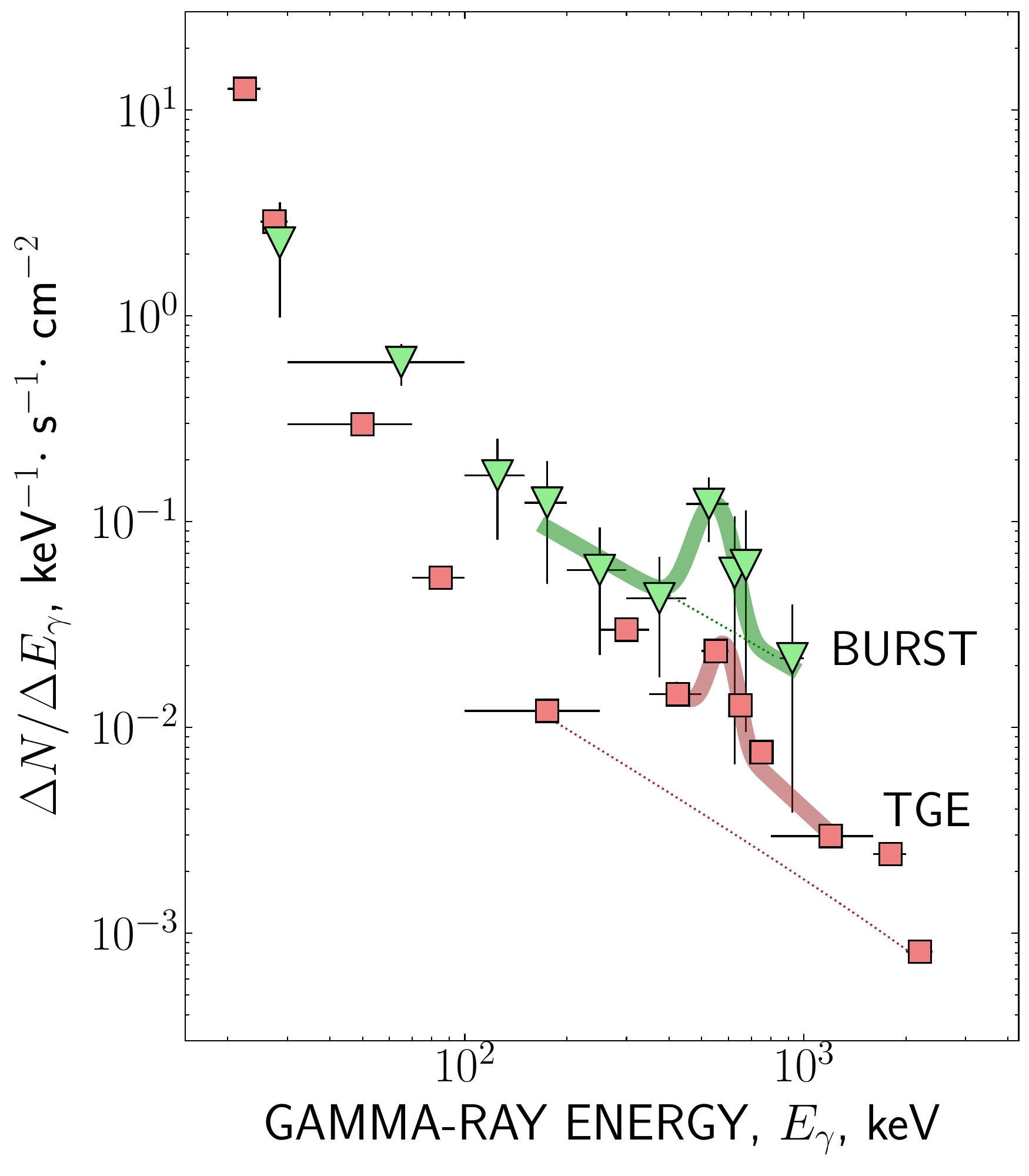}
\caption{Differential energy spectra of gamma radiation detected at the time of the TGE event (below, squares) and of the short-time burst (above, triangles). Background is subtracted. The thick continuous curves indicate the deposit from the annihilation line.}
\label{figidiffispc}
\end{figure*}

The question on presence of the positron component among radiations generated in thundercloud takes a special place by experimental investigations in the field of atmospheric electricity. Possible source of positrons at thunderstorm time can be either electromagnetic interaction of accelerated runaway~particles, as suggested by  \cite{thunderpositron_gurevich1999}, or $\beta^+$~decay of short living radioactive nuclei which arise under influence of gamma rays from the developing electron-photon avalanche due to the mechanism of photonuclear interaction, as it was discussed \textit{e.g.} in \cite{thunderpositron_enoto2017,thunderneutro_babich_2019}. Hence,  investigation of the energy and temporal characteristics of positron flux helps to shed light on various features of atmospheric electric discharge.

A convenient marker of positron production is observation of the $e^{\pm}$ annihilation line at 511\,keV in the spectrum of gamma emission. To disclose most prominently the presence of this line in the events considered in present publication the differential energy spectra $dI/dE_\gamma$ were built on the basis of integral data from Figure~\ref{figigammospc}. For this purpose it was used the next to last (\textit{-1s} labelled) integral spectrum on the left hand plot of this figure, which has just preceded the terminating discharge at the time of TGE event, and the spectrum  of the gamma ray burst in the right panel. As before, the differential spectra $dI/dE_\gamma$ in each point were normalized to the total sensitive area of the scintillation detector crystal, and to the efficiency of gamma radiation detection, as the latter is given by  Figure~\ref{figigammoeffici} for corresponding $E_\gamma$. Together with normalization, the levels of the background radiation intensity (those shown with circles in Figure~\ref{figigammospc}) were subtracted from both spectra. The differential energy spectra thus calculated are presented in Figure~\ref{figidiffispc}. In such form these spectra agree rather well, both by their shape and absolute intensity, with the differential spectrum of TGE emission reported by
\cite{tsuchia_2011_tgf_spectrum_from_dwyer}.

As it follows from Figure~\ref{figidiffispc}, a statistically significant peak can be found indeed within the energy range of (400--600)\,keV in gamma ray spectra of both the TGE event and of the short-time radiation burst.  An excess of both enhancements over the extrapolated levels of ``regular'' power spectrum shown with dotted lines is of about $\simeq$2\,s$^{-1}$cm$^{-2}$ for the TGE, and of $\simeq$10\,s$^{-1}$cm$^{-2}$ in the case of the burst (by latter estimations the width of the energy spectrum bins accepted at the measurements time, correspondingly that of 100\,keV and 150\,keV, was taken into account).
% tge: 2.3e-2 against 0.3e-2 = 2.e-2
%                             x 100keV ~2 s^{-1}cm^{-2}
% tgf: 1.2e-1 against 0.3e-1 = 0.9e-1
%                              x 150keV ~14 s^{-1}cm^{-2}
If to interpret the peaks as originated from annihilation line, these two values correspond to an absolute contribution of the 511\,keV gamma rays emitted due to $e^\pm$ reactions within a (50--100)\,m neighbourhood of discharge region. Since it cannot be excluded an admixture of other radiation sources into this spectrum range both estimates should better be meant in the sense of an upper possible limit.

Besides the supposed annihilation peak, the differential spectrum detected at the time of the TGE event demonstrates another irregularities and amplitude enhancements aro\-und the energy ranges of about $\sim$300\,keV and (1500--2000)\,keV. Similar features were reported in the works of \cite{thundertge_bogomolov2015} and \cite{thundertge_bogomolov2016} which were aimed to precision study of radiation spectra of TGEs observed, correspondingly, at the ground level (in Moscow region), and at Aragats mountain. In these publications it is claimed that such deviations of TGE spectra from uniform power behaviour correspond to the gamma ray lines originating from radioactive decays of $^{222}$Rn nuclei and their daughter products. Since there was a many hours long rainy period during the beginning of the day of August~12, 2018 which has immediately preceded the TGE, and the radon concentration in the near-earth atmosphere is known to increase considerably at precipitation time such explanation seems to be quite plausible in our case.

\subsection{Neutron signal detected at the time of thunderstorm activity}

\begin{figure*}
\centering
\includegraphics[width=0.49\textwidth, trim=5mm 0mm 0mm 0mm]{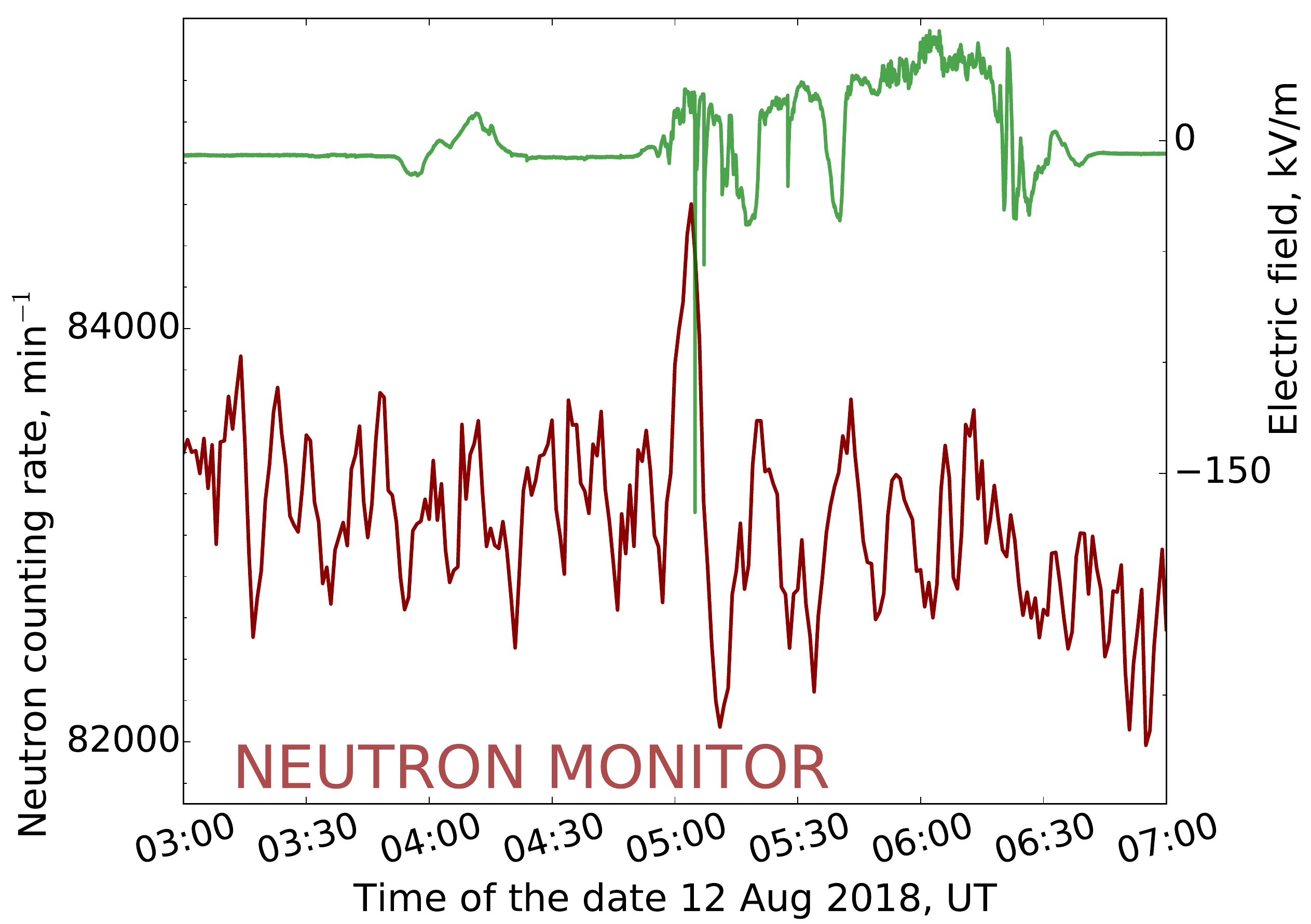}
\includegraphics[width=0.49\textwidth, trim=0mm 0mm 5mm 0mm]{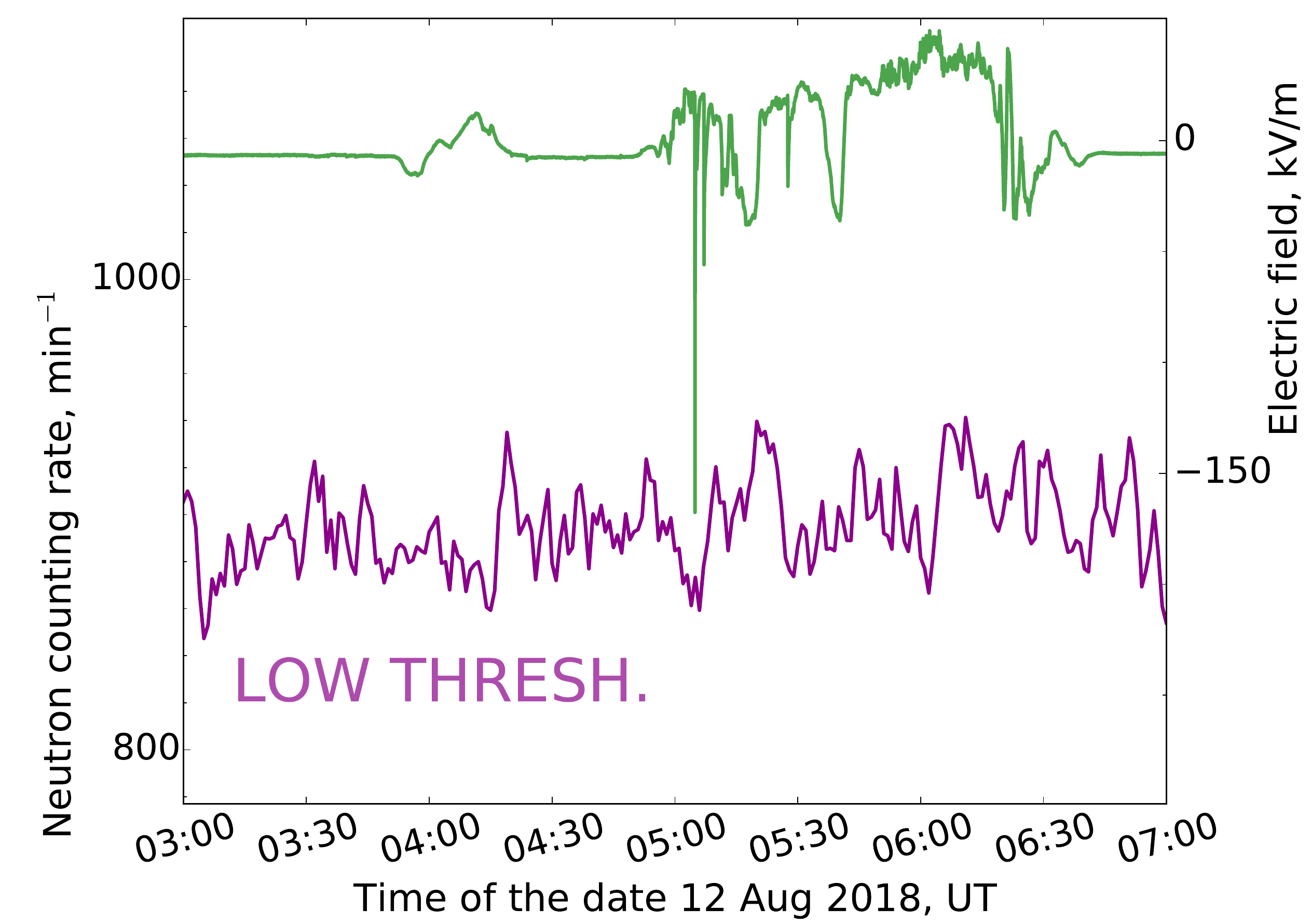}
%\\
%\includegraphics[width=0.49\textwidth, trim=5mm 0mm 0mm 0mm]{170613nm64.pdf}
%\includegraphics[width=0.49\textwidth, trim=0mm 0mm 5mm 0mm]{170613thermal.pdf}
%\caption{Time series of the counting rate of neutron signals in the Tien Shan neutron monitor, and in the low-threshold neutron detector at the time of the August 12, 2018 TGE (top row) and June 13, 2017 TGF (bottom row) events. Upper curve in all plots indicates the simultaneous measurements of local electric field.}
\caption{Time series of the counting rate of neutron signals in the Tien Shan neutron monitor, and in the low-threshold neutron detector at the time of the August 12, 2018 TGE event. Upper curve in the plots indicates simultaneous measurements of the local electric field.}
\label{figinm64}
\end{figure*}

Together with the data on gamma radiation, the measurements of the intensity of neutron background in thunderstorm time are available at Tien Shan station as well.  As it was explained in Instrumentation section, two types of neutron detectors are used for this purpose in Tien Shan experiments: the neutron monitor for detection of evaporation neutrons produced in interaction of high energy cosmic ray particles, and the low-threshold detector for registration of the environmental background of thermal neutrons. In particular, the counting rates of neutron signals were registered at the time of the TGE event considered above, and the results of these measurements are presented in Figure~\ref{figinm64}. Before to be plotted here the original neutron counts which have been recorded continuously every minute at both installations were corrected to variation of atmospheric pressure, and the moving average filter algorithm was applied to them to stress more distinctly any systematic intensity variation against the background of random fluctuations. The length of the filter kernel accepted in latter procedure was equivalent to a 3\,min long dataset, which is comparable with total duration of the TGE event.

As it is seen in the left panel of Figure~\ref{figinm64}, in spite of a rather rough time resolution of neutron data a prominent intensity excess does exist in the record of the neutron monitor signals made in the day of August 12, 2018, with its position being superimposed on duration of the observed TGE event. The relative amplitude of this transient increase above the background is of about 2\%, while its highest peak value just precedes the negative jump of the local electric field which was detected at the moment of TGE termination. Just after TGE a (0.5--1)\% deep depression is seen in the monitor intensity record. In contrary, simultaneous data on the thermal neutron background presented in the right plot window of Figure~\ref{figinm64} do not demonstrate any irregularities in TGE time which could exceed the level of usual statistical fluctuations.

An absolute amplitude of the peak excess of the neutron counting rate above its usual background $\mathcal{R}$ at the time of TGE occurs to be of about (1300--1400)\,min$^{-1}$, or, with account of the sum sensitive area of the neutron monitor (18\,m$^2$), $\mathcal{R}\approx 10^{-4}$\,s$^{-1}$cm$^{-2}$. Seemingly, this estimation can give a hint on the origin of this peak.

According to \cite{thunderneutro_babich_2019}, presently it is believed that the most probable source of neutron generation at thunderstorm times is connected with photonuclear interaction of the hard gamma rays with $E_\gamma \gtrsim 7$\,MeV. In the case if this interaction takes place somewhere outside the detector, \textit{e.\,g.} in atmosphere, or in the soil around the monitor, an excessive signal count in the latter must be caused by the flux of low-energy (thermal) neutron component, since the neutrons born in photonuclear reactions loose their initial MeV-order energy rather quickly on their way to the detector site. With account of the $\sim$1\% registration probability of thermal neutrons by NM64 type supermonitor (see Figure~\ref{figineutroeffici}), an agreement with the above $\mathcal{R}$ value can be achieved only if the neutron flux at the time of TGE was of about $I_n\sim 0.01$\,s$^{-1}$cm$^{-2}$. This value is an order of magnitude above the typical background level registered by the low-threshold neutron detector ($I_{bckgr}\approx 0.002$\,s$^{-1}$cm$^{-2}$), and consequently such an excess should be prominently seen  in the counting rate record of the latter, but according to the right plot of Figure~\ref{figinm64} this is not the case. Hence, the supposition on \textit{external} neutron generation in the space outside the monitor fails.

Alternatively, it is possible to suppose after \cite{tsuchia2014} the origin of neutrons from photonuclear reactions which take place not outside but immediately \textit{within} the monitor volume. In this variant the intensity of the $E_\gamma\gtrsim 7$\,MeV radiation must be of about $I_\gamma\sim 0.1$\,s$^{-1}$cm$^{-2}$, with account of $\mathcal{R}$ value and geometrical area of the monitor, as well as of the energy threshold and efficiency of neutron production by gamma ray quanta given \textit{e.\,g.} in \cite{thunderneutrotibet2012}. At the same time, an extrapolation into this energy range of the $f(E_\gamma)$ approximation function which was introduced by discussion of the left panel plot in Figure~\ref{figigammospc} results in estimation of $\simeq$0.003\,s$^{-1}$cm$^{-2}$ only, \textit{i.\,e.} it is two orders of magnitude below the necessary $I_\gamma$ value, even in the case of the most intensive spectrum which has immediately preceded the terminating discharge.

Hence, the photonuclear interaction can not explain the increase of the neutron monitor counting rate which was observed at the time of the considered TGE event.

Another possible mechanism of neutron generation just within the monitor is connected with the muonic component of cosmic rays, as it was discussed by \cite{muraki_on_muons_2004,alexeenko2004_onmuons,dormandorman_on_muminus_2005}, and primarily with the process of $\mu^-$ capture by atomic nuclei. According to the results of a Geant4 simulation made by \cite{our2018_undg}, the peak of the average neutron production multiplicity in monitor because of this effect is of about $\bar{\nu}\simeq (0.05-0.1)$ neutrons per an incident muon, and this maximum is reached in the muon energy range of $\sim$(100--500)\,MeV. With such $\bar{\nu}$, an additional $\sim$0.001\,s$^{-1}$cm$^{-2}$ order flux of negative muons is quite sufficient to explain the observed amplitude $\mathcal{R}$ of the neutron intensity increase. The latter estimation is a small value in comparison with the total background flux of  the $\gtrsim$100\,MeV muons at Tien Shan station (which is of about $\simeq$5\,s$^{-1}$cm$^{-2}$ according to the integral energy spectra presented in \cite{our2018_undg}), and such variation can be con\-ve\-ni\-ent\-ly accounted for by modulation influence of the local electric field. Indeed, as it is seen in Figure~\ref{figinm64}, in the time of the neutron intensity increase the near-earth field sensor detected a positive field of the order of $+$(25--30)\,kV/m. As it has been just noticed above by discussion of the TGE event, this field may be accounted for acceleration of  negatively charged particles, both electrons and muons, in downward direction to the earth surface. In such a case the growth of the neutron counting rate at the TGE time can be explained by favorable disposition of the different domains of electric field just above the neutron monitor, quite analogously to situation with the gamma ray and electron TGE components. Then, a deep gap in the record of the neutron monitor counting rate after TGE may be a consequence of $\mu^-$ deceleration caused by a strong opposed field which has appeared after its reversal in the moment of terminating discharge. Such observations demonstrate a very fine sensitivity of muon signal as a messenger on the structure of electric fields in thundercloud which effect can be used in further investigations of thunderstorm phenomena.

\begin{table*}
\caption{\label{tabdeposi}
Peak intensity (in the units of s$^{-1}$cm$^{-2}$) of the thunderstorm activity connected emissions which were detected at the time of two considered atmospheric discharge events.
}

\newcommand{\thisherecolumwith}{0.105\textwidth} % 8 columns
\begin{tabular}{*{7}{c}}
\hline

% zeroth
\parbox[c][7ex]{0.05\textwidth}{~}  %0
\parbox{\thisherecolumwith}{\centering $\gamma \gtrsim 30$\,keV}& %1
\parbox{\thisherecolumwith}{\centering $\gamma \gtrsim 300$\,keV}& %2
\parbox{\thisherecolumwith}{\centering $\gamma \gtrsim 1$\,MeV}& %3
\parbox{\thisherecolumwith}{\centering 511\,keV\\line}  %6
\parbox{\thisherecolumwith}{\centering $e^- \gtrsim 2$\,MeV}&  %4
\parbox{0.12\textwidth}{\centering $e^- \gtrsim 100$\,MeV}& %5
\parbox{\thisherecolumwith}{\centering $\mu^-$}  %7
\\
\hline

% first
\parbox[c][5ex]{0.05\textwidth}{TGE}  %0
\parbox{\thisherecolumwith}{\centering 30}& %1
\parbox{\thisherecolumwith}{\centering 15}& %2
\parbox{\thisherecolumwith}{\centering 5}& %3
\parbox{\thisherecolumwith}{\centering $\leqslant$2}  %6
\parbox{\thisherecolumwith}{\centering 1.3}&  %4
\parbox{0.12\textwidth}{\centering ---}& %5
\parbox{\thisherecolumwith}{\centering $\gtrsim$0.001}  %7
\\

% second
\parbox[c][5ex]{0.05\textwidth}{burst}  %0
\parbox{\thisherecolumwith}{\centering 100}& %1
\parbox{\thisherecolumwith}{\centering 50}& %2
\parbox{\thisherecolumwith}{\centering 20}& %3
\parbox{\thisherecolumwith}{\centering $\leqslant$10}  %6
\parbox{\thisherecolumwith}{\centering $\sim$2}&  %4
\parbox{0.12\textwidth}{\centering $\sim$1}& %5
\parbox{\thisherecolumwith}{\centering ---}  %7
\\

\hline
\end{tabular}
\end{table*}

% note: though the table above relates logically to Conclusion, it might be better to settle it BEFORE the corresponding '\section' command for adequate placement on the final page when trying to compile a two-column pdf version for 'arXiv' site.
\section{Conclusion}

Two different kinds of the hard radiation flashes connected with thunderstorm activity were detected at Tien Shan Mountain Cosmic Ray Station near the moments of close lightning discharges: a prolonged TGE type event preceding a lightning, and a short-time radiation burst emitted just in the moment of a lightning discharge. The measurements of radiation fluxes were made in immediate vicinity ($\lesssim$100\,m) to spatial region of their generation in thunderclouds.
Both events differ significantly by their general time profiles, absolute intensity, and energy of emitted radiation. In spite of this difference the signs were noticed in both cases among the energy spectra of detected gamma radiation of the presence of 511\,keV annihilation line, and consequently of possible positron production in thunderclouds.

In the case of TGE event it was detected a statistically significant increase of the counting rate in the neutron monitor of Tien Shan Mountain Station. Seemingly, the origin of the surplus neutron production in the monitor can be attributed to modulation influence of thundercloud electric field on the flux of negative cosmic ray muons. %If such an effect would be confirmed by further observations, detection of thin variations in the counting rate of neutron monitor could be a sensitive instrument of the large scale structure of electric fields in thundercloud.

The absolute intensity estimations which were made all over the article for  contribution on the part of thunderstorm connected emission into the various components of detected radiation are summed up together in Table~\ref{tabdeposi}.

\section*{Acknowledgement}
This work was supported by ``Applied Space Research'' program, the project \#0118RK00802; by the grant \#BR053\-363\-83 of Aerospace Committee of Ministry of Digital Development, Innovation, and Aerospace Industry of Republic of Kazakhstan; and by the grants \#BR052\-362\-91 and \#BR052\-364\-94 of the program ``Fundamental and applied research in related fields of physics of terrestrial, near-Earth and atmospheric processes, and their practical application'' administered by Ministry of Science and Education of Republic of Kazakhstan.

%\section*{References}
% the explicit header is to be commented out - sic!

\end{document}